\documentclass[10pt,twocolumn,letterpaper]{article}
\ifdefined\pdfsuppressptexinfo
\fi

\usepackage[pagenumbers]{cvpr} 

\usepackage{amsmath,amsfonts,bm}

\newcommand{\sys}{\textsc{Tailor}}
\newcommand{\sysfull}{\textsc{Tailor-F}}
\newcommand{\sysgreedy}{\textsc{Tailor-G}}

\def\eqref#1{equation~\ref{#1}}

\def\1{\bm{1}}

\DeclareMathAlphabet{\mathsfit}{\encodingdefault}{\sfdefault}{m}{sl}
\SetMathAlphabet{\mathsfit}{bold}{\encodingdefault}{\sfdefault}{bx}{n}

\usepackage{adjustbox}
\usepackage{xurl}
\usepackage{multirow}
\usepackage{array,tabularx}
\usepackage{colortbl}
\usepackage{amsthm}
\usepackage{tablefootnote} 
\usepackage{microtype}
\newcommand{\papertablestyle}{%
  \small
  \setlength{\tabcolsep}{4pt}%
  \renewcommand{\arraystretch}{1.08}%
}
\newcommand{\papertablenote}[1]{%
  \par\smallskip
  \begin{minipage}{\linewidth}\centering\footnotesize
  #1
  \end{minipage}%
}

\newtheorem{proposition}{Proposition}

\newtheorem{corollary}{Corollary}

\definecolor{hl}{RGB}{225,240,255}          

\newif\ifshowrev \showrevtrue
\newcommand{\revcolor}{\ifshowrev\color{red}\fi}
\DeclareRobustCommand{\rev}[1]{{\revcolor #1}}            
\newenvironment{revblock}{\par\revcolor}{\par}              

\newcommand{\paraheadskip}{.5em}
\renewcommand{\paragraph}[1]{\par\vspace{\paraheadskip}\noindent\textbf{#1}}

\newcommand{\fitdisplayskips}{}
\appto\normalsize{\fitdisplayskips}
\newcommand{\fiton}{%
  \setlength{\abovecaptionskip}{.5em}
  \setlength{\textfloatsep}{8pt plus 2pt minus 4pt}
  \setlength{\dbltextfloatsep}{8pt plus 2pt minus 4pt}
  \renewcommand{\paraheadskip}{0pt}
  \setlist[itemize]{noitemsep,topsep=0em,wide=0pt}
  \renewcommand{\fitdisplayskips}{
    \abovedisplayskip=5pt plus 2pt minus 3pt\relax
    \belowdisplayskip=\abovedisplayskip
    \abovedisplayshortskip=0pt plus 2pt\relax
    \belowdisplayshortskip=4pt plus 2pt minus 2pt\relax}%
  \normalsize}
\newcommand{\fitoff}{%
  \setlength{\abovecaptionskip}{10pt}%
  \setlength{\textfloatsep}{20pt plus 2pt minus 4pt}%
  \setlength{\dbltextfloatsep}{20pt plus 2pt minus 4pt}%
  \renewcommand{\paraheadskip}{.5em}%
  \setlist[itemize]{noitemsep,leftmargin=*,topsep=0em}%
  \renewcommand{\fitdisplayskips}{}%
  \normalsize}

\definecolor{cvprblue}{rgb}{0.21,0.49,0.74}
\usepackage[pagebackref,breaklinks,colorlinks,allcolors=cvprblue]{hyperref}

\def\confName{CVPR}
\def\confYear{2027}

\title{Made to Measure: Designing Image Watermarks to Specification}

\author{
Mingzhe Li$^{1}$ \quad
Yuefeng Peng$^{1}$ \quad
Kejing Xia$^{2}$ \quad
Pranav Jeyakumar$^{1}$ \\
Ruolan Leslie Famularo$^{3}$ \quad
Shiqing Ma$^{1}$ \\[0.5em]
$^{1}$University of Massachusetts Amherst \quad
$^{2}$Georgia Institute of Technology \quad
$^{3}$Dolby Laboratories
}

\begin{document}
\fiton
\maketitle
\AddToHook{para/begin}[tailor-fit]{\looseness=-1}

\begin{abstract}
Image watermarking enables provenance and attribution, but practical
deployments must jointly satisfy robustness, false-positive rate
(FPR), image quality, and latency requirements. Existing watermarking
methods offer complementary robustness against different attacks,
making their composition attractive for broader protection.
However, combining watermarks introduces interference, distortion,
decoding overhead, and additional verification paths that share
the same FPR budget.
We propose \sys{}, a request-conditioned watermark composition
framework with three stages: (1) \emph{offline characterization}
models fragment recovery, distortion, latency, and composition
interactions across embedding strengths; (2) \emph{joint configuration
selection} uses satisfiability modulo theories (SMT) to jointly
optimize fragment selection, continuous embedding strengths,
embedding order, and geometric recovery while minimizing predicted
distortion; and (3) \emph{live calibration} validates candidate
configurations on user images and refines inaccurate offline
predictions to improve deployment success.
Across 7,321 distinct deployment requests spanning five scenarios
and 20 attack settings, \sys{} achieves 96.21\% scenario-averaged
request satisfaction with a mean PSNR of 41.02~dB, demonstrating
its effectiveness in meeting diverse deployment requirements
while maintaining high image fidelity.
Code: \url{https://github.com/aaFrostnova/Tailor}.
\end{abstract}

\section{Introduction}
\label{sec:intro}

The growing use of advanced generative image models such as
\textit{GPT-Image-2}~\citep{openai2026gptimage2},
and \textit{Nano Banana 2}~\citep{google2026nanobanana2}
has made provenance and attribution important concerns for creators and
service providers~\cite{li2026promptminer,li2025editor}. Invisible image watermarking associates an image with a machine-readable signal that can be checked after distribution~\citep{trustmark,vine}, an approach that has already seen commercial deployment and demonstrated significant practical value. For example, Google introduced SynthID for Imagen customers on Vertex AI to help identify AI-generated images~\citep{google_synthid2023}, while Meta introduced invisible watermarking for images generated by Meta AI to improve transparency and traceability~\citep{meta_invisible_watermark2023}. These deployments demonstrate the practical role of watermarking in provenance and attribution.

\begin{figure*}[t]
\centering
\includegraphics[width=0.90\linewidth]{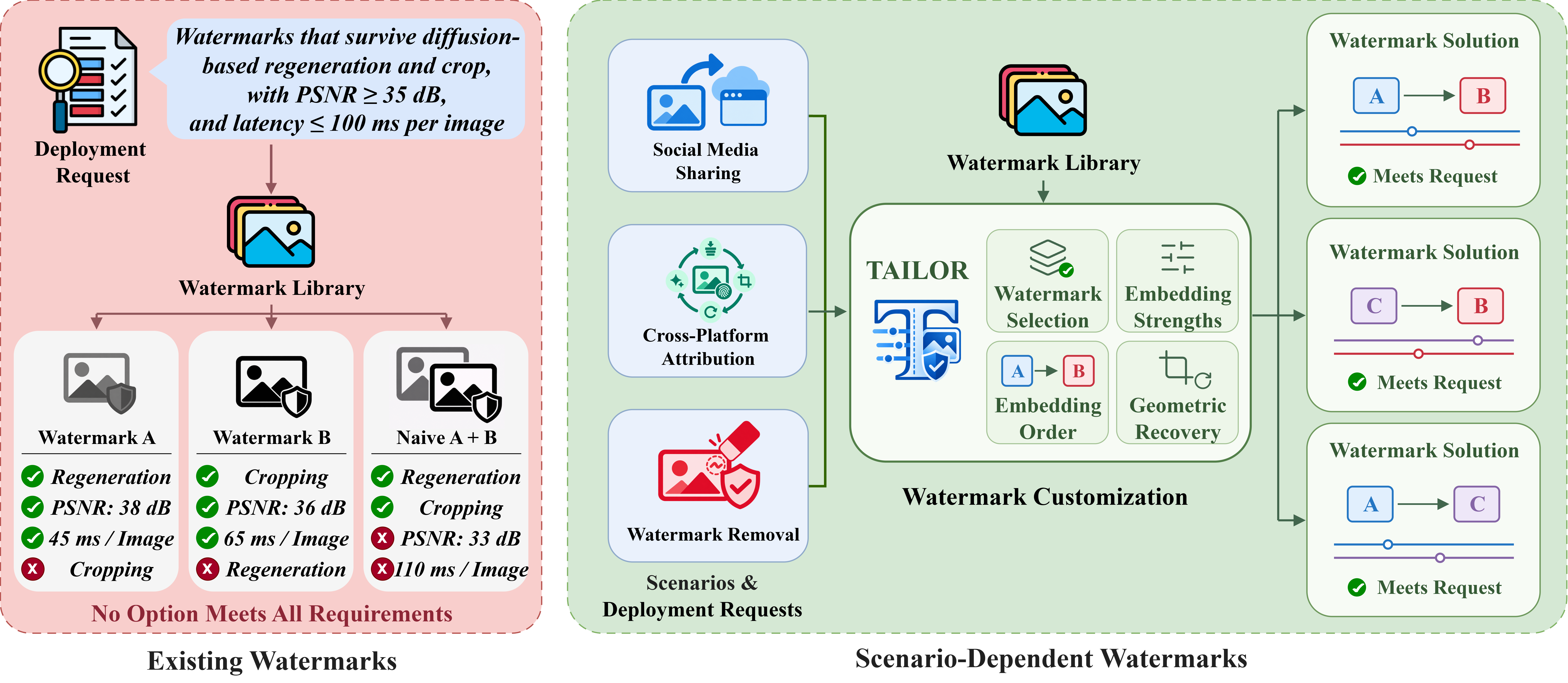}
\caption{\textbf{Illustration of TAILOR.} Existing watermark schemes may fail to jointly satisfy scenario-specific robustness, quality, and latency requirements, while TAILOR composes and configures complementary watermarks to produce deployment-specific solutions.}
\label{fig:intro_illustration}
\end{figure*}

Existing watermarking methods improve robustness through different designs. TrustMark learns to balance image quality and watermark recovery under image perturbations~\citep{trustmark}. VINE leverages generative priors and surrogate distortions to improve robustness against image editing~\citep{vine}, while VideoSeal targets efficient watermarking with robustness to geometric transformations and video compression~\citep{videoseal}. Their designs therefore provide complementary protection against different transformations, while selecting a single watermark or tuning its strength cannot exploit the protection offered by the others. Combining multiple watermark mechanisms is thus attractive for achieving broader robustness.

Practical deployment, however, requires more than robustness alone. A watermark configuration must survive expected transformations while preserving image quality, satisfying a false-positive-rate (FPR) requirement, and meeting runtime constraints. For example, SynthID is designed to remain detectable after common image modifications while preserving image quality~\citep{google_synthid2023}, and the SynthID-Image report further highlights low encoding overhead, high-throughput decoding, and reliable decisions at low FPRs for large-scale deployment~\citep{gowal2025synthid}. Beyond watermark detection, C2PA provides a standardized framework for cryptographically verifiable provenance across content workflows~\citep{c2pa}, while Adobe Content Authenticity combines invisible watermarking with Content Credentials and digital fingerprinting to preserve attribution after transformations such as metadata removal and screenshots~\citep{adobe_contentauth2024,adobe_contentauth_impact2025}. Deployment and access conditions can further change the relevant attack surface~\citep{gowal2025synthid}. When multiple watermarks are combined, their fragments must share the same quality, latency, and FPR budgets, while their embedding strengths and order jointly affect robustness and distortion. Moreover, offline measurements may not transfer directly to the user's images. These coupled requirements motivate treating the expected attacks, FPR budget, image-quality floor, and runtime limit as deployment-specific inputs. The challenge is therefore to jointly select and validate a watermark composition that satisfies the complete request.

To address these challenges, we propose \sys{}, a request-conditioned framework for composing and validating image watermarks (\autoref{fig:intro_illustration}). The core insight is to treat existing watermark encoder and decoder pairs as complementary \emph{fragments} whose configurations should be optimized jointly rather than selected independently. \sys{} first characterizes fragment recovery, distortion, runtime, and geometric recovery offline and represents these measurements as continuous response models. Given a request specifying attacks, FPR, image quality, and latency, an SMT-based solver~\citep{barrett2018satisfiability} jointly selects the fragments, embedding strengths, order, and geometric recovery stages that satisfy the complete specification while minimizing predicted distortion. Live calibration then evaluates candidate compositions on the user's images and refines inaccurate offline predictions when necessary. 
We conduct a comprehensive evaluation of \sys{} on 7,321 distinct deployment requests spanning five scenarios and 20 attack settings. \sys{} achieves 96.21\% scenario-averaged request satisfaction with a mean PSNR of 41.02~dB, outperforming existing methods in
robustness while achieving consistently better image quality (\autoref{tab:main}). Controlled ablations further validate the contributions of candidate search, watermark composition, and strength adaptation. Our contributions are summarized as follows:

\begin{itemize}
    \item We formulate watermark deployment as a
    request-conditioned composition problem.
    Rather than evaluating watermarks in isolation,
    we consider deployment requirements that jointly
    specify attack coverage, an FPR budget, image
    quality, and latency constraints.

    \item We propose \sys{}, a measurement-guided
    framework that jointly optimizes watermark
    selection, embedding strengths, embedding order,
    and geometric recovery using SMT.
    Live calibration validates candidate configurations
    on user images and refines performance predictions
    to improve deployment success.

    \item We evaluate \sys{} on 7,321 deployment
    requests across five scenarios and 20 attack
    settings. \sys{} achieves 96.21\% scenario-averaged
    request satisfaction, outperforming Optimized
    Single (73.37\%), Always All Three (94.47\%),
    and Greedy (91.05\%). Ablation studies further
    validate the benefits of watermark composition,
    strength optimization, and live refinement.
\end{itemize}

\section{Related Work}
\label{sec:related}

\paragraph{Image Watermarking and Attacks.}
Post-hoc watermarking uses transform-domain methods based on DWT with DCT~\citep{al2007combined} and DWT with DCT and SVD~\citep{navas2008dwt}, or learned encoder and decoder models such as RivaGAN~\citep{rivagan}, StegaStamp~\citep{stegastamp}, WAM~\citep{wam}, TrustMark~\citep{trustmark}, and VINE~\citep{vine}. These methods face signal-processing attacks such as JPEG compression, blur, and noise; geometric attacks such as cropping, resizing, and rotation~\citep{waves}; adaptive and optimization-based removal attacks such as WEvade~\citep{jiang2023evading} and UnMarker~\citep{unmarker}; and regeneration attacks such as Regen~\citep{zhao2024regen} and CtrlRegen$+$~\citep{ctrlregen}. Regeneration attacks reconstruct perceptually similar images while suppressing watermark signals~\citep{zhao2024regen,ctrlregen}. RAVEN~\citep{raven} further exploits diffusion-based novel-view synthesis for watermark removal. Different watermarks are effective against different attacks, while no single method is uniformly robust across the evaluated threat landscape~\citep{waves, zhang2026disappearing}. In our evaluation, VINE better withstands regeneration but is vulnerable to the tested geometric attacks, whereas TrustMark shows the opposite pattern. This complementarity motivates combining watermarks, although sequential embedding can introduce interference and image distortion, and additional decoder executions increase latency.

\paragraph{Scenario-Dependent Watermark Configuration.}
Practical deployments impose requirements beyond average watermark recovery. Adobe Content Authenticity combines invisible watermarking with Content Credentials and digital fingerprinting for provenance across transformations such as metadata removal and screenshots~\citep{adobe_contentauth2024,adobe_contentauth_impact2025}, while SynthID emphasizes image-quality preservation, robustness to common edits, and efficient detection~\citep{google_synthid2023,gowal2025synthid}. OmniGuard~\citep{omniguard} similarly combines copyright protection with manipulation localization, illustrating that watermarking objectives vary by application. FLEXMark~\citep{flexmark} further argues for evaluation under shared distortion budgets, realistic payloads, low false-positive operating points, and adaptive removal rather than average recovery alone. These works motivate configuration at the system level: a social-media deployment may prioritize crop and compression robustness, while another may prioritize regeneration resistance, attribution reliability, payload capacity, image quality, or latency.
Existing work develops stronger individual watermarks, broader attack benchmarks, fixed multi-purpose architectures, and watermark ensembles~\citep{petrov2025coexistence}, but to our knowledge does not directly treat deployment as a per-request constrained optimization over heterogeneous watermark fragments, their order, strengths, and recovery mechanisms. We therefore propose \sys{}, a request-conditioned framework that jointly selects watermark fragments, embedding order, strengths, and geometric recovery support under user-specified robustness, false-positive, image-quality, and latency constraints.

\section{Threat Model}

We consider two parties: a protector, who deploys and
verifies watermarks, and an attacker, who attempts to
remove watermark information from released images.

\paragraph{Protector's Goal and Capabilities.}
The protector specifies a deployment request
$u=(\mathcal A_u,\alpha,q,t)$, where $\mathcal A_u$
is the set of attacks the watermark is expected to
withstand, $\alpha$ is the per-image false-positive-rate
(FPR) budget for verifying a queried watermark identity,
$q$ is the minimum image-quality requirement, and $t$
is the deployment latency ceiling.
A false positive occurs when the verifier accepts an
image that does not contain the queried watermark
identity. The FPR budget $\alpha$ applies to the
complete verification of a single image against that
identity, including all verification paths enabled by
the deployed configuration.
Let $x_o$ denote an original image and
$w\in\{0,1\}^n$ denote the watermark payload
associated with an identity.
A watermark encoder $E(\cdot,\cdot)$ embeds $w$
into $x_o$ to produce $x_w=E(x_o,w)$.
An attack $a\in\mathcal A_u$ transforms the released
watermarked image into
$\tilde{x}_w^a=a(x_w)$.
The protector verifies the attacked image by comparing
the decoder readouts with the expected payload $w$.
The goal is to preserve reliable watermark verification
under the requested attacks while satisfying the FPR,
image-quality, and latency requirements in $u$.

We focus on post-hoc watermarking: the protector
may modify an image after it has been generated or
acquired, but does not modify the underlying image
generator or its training procedure.
Verification is reference-free: given a candidate
image, a queried identity, the secret key, and the
deployed configuration, the verifier determines
whether the image carries that identity without
requiring the original unwatermarked image.

\paragraph{Attacker's Goal and Capabilities.}
The attacker aims to transform the released
watermarked image $x_w$ so that its embedded identity
can no longer be reliably recovered.
We assume the attacker has access to $x_w$, but
not the protector's secret key or the original
unwatermarked image $x_o$.
Depending on the requested attack, the attacker may
apply signal-processing or re-encoding operations,
geometric transformations, neural compression,
publicly available regeneration or editing models,
or optimization-based watermark-removal attacks.
For each request, the attacker is granted the
capabilities required by the attacks in $\mathcal A_u$.
Attacks are evaluated separately unless an attack
is explicitly defined as a composition of multiple
operations. Robustness is evaluated against the
attack set and settings specified by each request.

\section{Methodology}
\label{sec:method}

\begin{figure*}[t]
\centering
\includegraphics[width=1\linewidth]{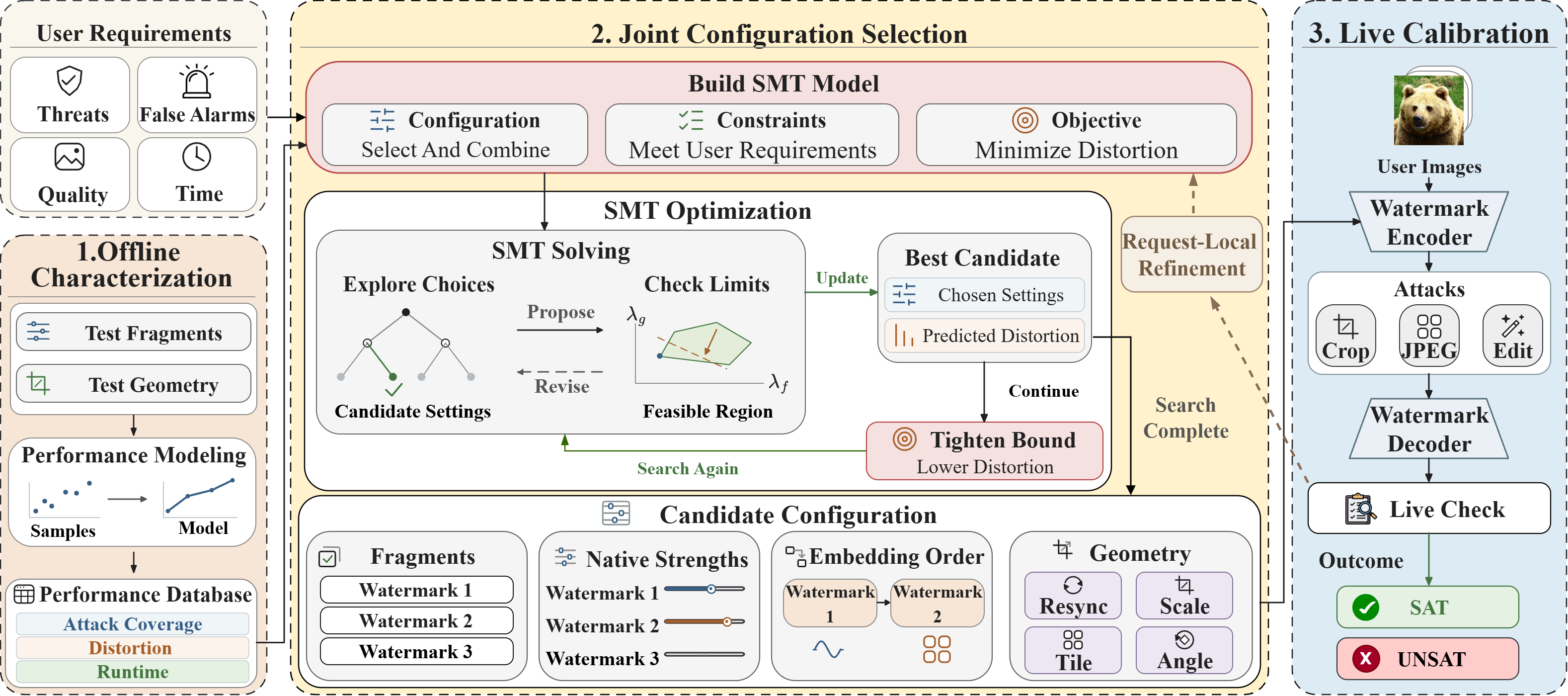}
\caption{\textbf{Overview of TAILOR.} (1) Offline characterization builds a performance database of attack coverage, distortion, and runtime. (2) Joint configuration selection solves an SMT model of the user requirements to find the lowest-distortion configuration of fragments, strengths, order, and geometry. (3) Live calibration validates this configuration on user images under the requested attacks, returning SAT or UNSAT.}
\label{fig:method-overview}
\end{figure*}

In this section, we introduce \sys{}, a request-conditioned watermark
composition framework. Given a deployment request
$u=(\mathcal A_u,\alpha,q,t)$,
\sys~ searches for a watermark composition satisfying
the specified attack coverage, false-positive-rate (FPR)
budget, image-quality floor, and latency ceiling.
The framework consists of three stages.
First, \textbf{Offline Characterization} profiles watermark
fragments and geometric recovery stages, storing their
measured performance in a database.
Second, \textbf{Joint Configuration Selection} formulates
the deployment request as an SMT optimization problem
over the performance database and searches for a feasible
configuration with minimum predicted distortion.
Third, \textbf{Live Calibration} evaluates candidate
configurations on the user's images under the requested
attacks and refines inaccurate performance predictions
when necessary.
Figure~\ref{fig:method-overview} provides an overview of \sys{}.

\subsection{Offline Characterization}
\label{sec:method-offline}

\paragraph{Measurements.}
We build a Performance Database by profiling three
watermark fragments,
$\mathcal{F}=\{\text{VINE~\cite{vine}},
               \text{TrustMark~\cite{trustmark}},
               \text{VideoSeal~\cite{videoseal}}\}$,
over their native embedding strengths and evaluated attacks.

We represent a candidate configuration as
$c=(S,\lambda,\pi,\varphi)$, where
$S\subseteq\mathcal{F}$ is the selected fragment subset,
$\lambda=(\lambda_f)_{f\in S}$ specifies the native
embedding strengths, $\pi=(f_1,\ldots,f_{|S|})$
denotes the embedding order, and $\varphi$ is an
optional geometric recovery stage.
We use $f,g\in\mathcal{F}$ to denote watermark
fragments and $a\in\mathcal A_u$ to denote a requested
attack. The database contains five types of measurements:

\begin{itemize}
    \item \textbf{\textit{Fragment Recovery.}}
    For each fragment $f$, strength $\lambda_f$, and
    attack $a$, we measure mean bit accuracy
    $b_f(\lambda_f,a)$ and retain the underlying
    per-image scores.
    These measurements predict attack coverage and
    support image-level acceptance estimates under
    the configuration-dependent verification threshold
    $\tau(c,\alpha)$, where $\alpha$ is the FPR budget.
    Let $\rho$ denote the required image-acceptance rate.
    The indicator $\Gamma_{f,a}(c)$ specifies whether
    fragment $f$ is predicted to satisfy the screening
    acceptance-rate requirement under attack $a$.

    \item \textbf{\textit{Embedding Distortion.}}
    We measure standalone embedding distortion
    $D_f(\lambda_f)$ as mean squared error (MSE)
    between the original and watermarked images.
    These measurements support the prediction of
    total embedding distortion $D(c)$ and its
    corresponding image quality $P(c)$ in PSNR.

    \item \textbf{\textit{Runtime.}}
    We measure the embedding and decoding latency
    of each fragment.
    These measurements, together with geometric-stage
    overhead, support the prediction of total deployment
    latency $L(c)$.

    \item \textbf{\textit{Composition Interactions.}}
    We measure the recovery loss
    $\delta_{g\rightarrow f}(\lambda_g,a)$ of fragment
    $f$ when fragment $g$ is embedded after it, and
    the excess pairwise distortion
    $e_{fg}(\lambda_f+\lambda_g)$ beyond their
    standalone MSE terms.
    These measurements capture the effects of
    fragment composition and embedding order
    on recovery and distortion.

    \item \textbf{\textit{Geometric Recovery.}}
    We profile four optional geometric recovery stages
    $\varphi$: scale, tile, resynchronization, and
    angle search (Appendix~C).
    For each stage, we measure its recovery performance,
    decoding overhead, and additional embedding
    distortion $D_\varphi(c)$ when applicable.
    Scale and tile recover VINE and TrustMark,
    respectively, from candidate crops, while
    resynchronization and angle search restore the
    image frame and decode the selected fragments.
    These measurements characterize the effects
    of geometric recovery on robustness, quality,
    and latency.
\end{itemize}

All offline measurements use fixed image subsets
disjoint from the images used for live calibration
(Appendix~D.1).

\paragraph{\textit{Continuous Strengths.}}
The Performance Database records strength-dependent
measurements at discrete values, while configuration
selection optimizes embedding strengths continuously.
We therefore represent each measured response using
piecewise-linear (PWL) interpolation.
Given measurements $y^{(k)}$ at knots
$\lambda^{(0)}<\cdots<\lambda^{(K)}$, we define
\begin{equation}
\begin{aligned}
\widehat y(\lambda)
&=y^{(k)}+s^{(k)}
\big(\lambda-\lambda^{(k)}\big),
\qquad
\lambda\in\big[\lambda^{(k)},\lambda^{(k+1)}\big],\\
s^{(k)}
&=\frac{y^{(k+1)}-y^{(k)}}
{\lambda^{(k+1)}-\lambda^{(k)}}.
\end{aligned}
\label{eq:pwl}
\end{equation}
Here, $y$ represents a generic strength-dependent
measurement. The resulting curves reproduce the
measured values at the knots and interpolate within
the characterized strength range.
In the following formulation, strength-dependent
performance functions refer to their corresponding
PWL-interpolated responses.
Each segment can be encoded in SMT using Boolean
segment-selection variables and linear arithmetic
constraints (Appendix~\ref{sec:method-preliminaries}).

\paragraph{\textit{Predicted Quantities.}}
Using the Performance Database, we predict the recovery,
embedding distortion, and latency of a candidate
configuration $c$.

\begin{itemize}
    \item \textbf{\textit{Recovery.}}
    For each selected fragment $f$ and attack $a$,
    we combine its standalone response with the
    measured recovery losses caused by subsequently
    embedded fragments:
    \begin{equation}
    \widehat b_f(c,a)
    =
    b_f(\lambda_f,a)
    -
    \sum_{g\in S:\,f\prec_\pi g}
    \delta_{g\to f}(\lambda_g,a),
    \label{eq:recovery}
    \end{equation}
    where $f\prec_\pi g$ indicates that $f$ is embedded
    before $g$. When the selected geometric recovery
    stage applies to fragment $f$ under attack $a$,
    we use the corresponding geometry-conditioned
    response measurements.

    \item \textbf{\textit{Embedding Distortion.}}
    We predict total embedding distortion in MSE by
    combining standalone distortion, measured pairwise
    corrections, and the contribution of the selected
    geometric stage:
    \begin{equation}
    D(c)
    =
    \sum_{f\in S}D_f(\lambda_f)
    +
    \sum_{\{f,g\}\subseteq S}e_{fg}(\lambda_f+\lambda_g)
    +
    D_\varphi(c),
    \label{eq:distortion}
    \end{equation}
    where $D(c)$ is the predicted total embedding MSE.
    Modeling distortion in MSE allows the PWL responses
    to be encoded using linear arithmetic constraints.
    The corresponding predicted PSNR is
    $P(c)=10\log_{10}\big(255^2/D(c)\big)$.
    Therefore, the requested PSNR floor $P(c)\ge q$
    is equivalent to the distortion constraint
    $D(c)\le255^2\,10^{-q/10}$.
    Minimizing $D(c)$ is thus equivalent to maximizing
    predicted PSNR.

    \item \textbf{\textit{Latency and Acceptance.}}
    Predicted latency $L(c)$ combines the embedding
    and decoding costs of the selected fragments
    with the overhead of the selected geometric stage.
    The stored per-image scores estimate the fraction
    of images passing the configuration-dependent
    verification threshold.
    The corresponding screening indicator
    $\Gamma_{f,a}(c)$ is used alongside predicted
    mean bit accuracy to determine attack coverage
    during configuration selection
    (Sec.~\ref{sec:method-solver}).
\end{itemize}

Additional database construction details are provided
in Appendix~\ref{sec:method-table-details}.

\subsection{Joint Configuration Selection}
\label{sec:method-solver}

Given a deployment request
$u=(\mathcal A_u,\alpha,q,t)$,
\sys{} jointly selects watermark fragments, their
embedding strengths and order, and the geometric
recovery stage.
These decisions are coupled: adding a fragment may
improve attack coverage while increasing distortion,
latency, and the number of verification paths sharing
the FPR budget.
We therefore formulate configuration selection as
a joint constrained optimization problem.

\paragraph{Configuration.}
For a configuration $c=(S,\lambda,\pi,\varphi)$,
the composite encoder $E_c$ applies the selected
fragments sequentially:
\begin{equation}
\begin{aligned}
x^{(0)}&=x_o,\\
x^{(j)}&=E^{\varphi}_{f_j}
\big(x^{(j-1)},w;\lambda_{f_j}\big),
\quad j=1,\ldots,|S|.
\end{aligned}
\label{eq:composition}
\end{equation}
The configuration also determines a decoder $D_c$
with individual-fragment, fused, and geometric
verification paths.
All selected fragments carry the same 100-bit payload
$w$ after fragment-specific key-derived permutations
and sign masks, providing multiple recovery paths
for a common identity.

Joint configuration selection involves both discrete
decisions over fragment selection, embedding order,
and geometric recovery, and continuous decisions
over embedding strengths.
These decisions are tightly coupled: the selected
fragments and their order determine the recovery
and distortion models, while embedding strengths
jointly affect attack coverage and image quality.
Moreover, fragment selection and geometric recovery
determine the verification paths sharing the FPR budget.
The PWL response models in Eq.~\ref{eq:pwl} allow
strength-dependent performance constraints to be
expressed using linear real arithmetic, while discrete
configuration choices and active PWL segments are
represented by Boolean variables.
This formulation enables satisfiability modulo
theories (SMT) to jointly optimize fragment composition,
embedding order, geometric recovery, and continuous
embedding strengths under the deployment constraints.

\paragraph{Constraints.}
We encode the deployment requirements as constraints
on configuration $c$.

\begin{itemize}
    \item \textbf{\emph{FPR constraint.}}
    We define the FPR budget $\alpha$ for a single
    verification of one image against a queried watermark
    identity. A false positive occurs when the verifier
    accepts an image that does not contain that identity.
    For each image, a configuration may evaluate multiple
    verification paths, including individual-fragment,
    fused, and geometric readouts. Let $T(c)$ denote the
    maximum number of permitted readout tests for one
    image verification.
    Each readout produces a 100-bit matching score
    against the expected watermark payload. Under the
    null model, the probability that a readout matches
    at least $h$ bits is upper-bounded by
    $\Pr[B\ge h]$, where
    $B\sim\mathrm{Bin}(100,1/2)$
    (Appendix~\ref{sec:method-budget-details}).
    We allocate the per-image FPR budget $\alpha$
    across the $T(c)$ verification tests and define
    the normalized detection threshold:
    \begin{equation}
    \begin{aligned}
    \tau(c,\alpha)=\frac{1}{100}\min\Big\{
    &h\in\{0,\ldots,101\}:\\
    &\Pr[B\ge h]\le\frac{\alpha}{T(c)}
    \Big\}.
    \end{aligned}
    \label{eq:fpr-threshold}
    \end{equation}
    An image is accepted if any verification path
    reaches this threshold. By the union bound,
    the probability of falsely accepting a single
    image is at most $\alpha$ under the null model.
    Thus, configurations with more verification paths
    require stricter detection thresholds. The budget
    $T(c)$ counts the verification paths for one image,
    independently of the number of images evaluated
    during calibration or deployment.
    Appendix~\ref{sec:fpr-selection} establishes the
    corresponding bound after request-local configuration
    selection under its stated assumptions.

    \item \textbf{\emph{Attack coverage.}}
    We use screening margins to account for differences
    between offline performance predictions and live
    measurements.
    The mean-accuracy margin $\eta$ defines
    $\bar\tau(c,\alpha)=\min(1,\tau(c,\alpha)+\eta)$,
    while the rate margin $\eta_{\mathrm{rate}}$ defines
    $\bar\rho=\min(1,\rho+\eta_{\mathrm{rate}})$.
    Using the image-level acceptance indicator
    $\Gamma_{f,a}(c)$ introduced in Sec.~4.1, we define
    \begin{equation}
    \begin{aligned}
    \mathrm{Cov}_{f,a}(c)&\equiv
    \left[
    \widehat b_f(c,a)\ge\bar\tau(c,\alpha)
    \ \wedge\ \Gamma_{f,a}(c)
    \right],\\
    \mathrm{Cov}_a(c)&\equiv
    \bigvee_{f\in S}\mathrm{Cov}_{f,a}(c).
    \end{aligned}
    \label{eq:cov}
    \end{equation}
    Each requested attack must be covered by at least
    one selected fragment, allowing different fragments
    to provide complementary protection.
    The screening margins apply only during offline
    selection; live calibration uses the original
    threshold $\tau(c,\alpha)$ and acceptance rate $\rho$.

    \item \textbf{\emph{Quality and latency.}}
    The predicted image quality and deployment latency
    must satisfy $P(c)\ge q$ and $L(c)\le t$,
    respectively.
    The PSNR floor is encoded in SMT through the
    equivalent distortion bound derived in Sec.~4.1.
\end{itemize}

\paragraph{Objective.}
We seek the feasible configuration with minimum
predicted embedding distortion:
\begin{equation}
\begin{aligned}
\min_c\ D(c)
&\quad\text{s.t.}\quad \Phi_u(c),\\
\Phi_u(c)&\equiv
\mathrm{Struct}_u(c)\wedge
\bigwedge_{a\in\mathcal A_u}\mathrm{Cov}_a(c)\\
&\qquad\wedge
(P(c)\ge q)\wedge(L(c)\le t),
\end{aligned}
\label{eq:selection-objective}
\end{equation}
where $\mathrm{Struct}_u(c)$ collects configuration
validity and compatibility constraints, including
that every selected fragment contributes to at least
one requested attack.

\paragraph{SMT Optimization.}
We encode the configuration decisions and their
associated performance constraints as Boolean
formulas and linear real arithmetic constraints,
using the PWL representation described above.

We minimize predicted distortion $D(c)$ through
iterative satisfiability queries.
Starting from a feasible incumbent with distortion
$D_{\mathrm{best}}$, the solver checks whether
\begin{equation}
\Phi_u(c)\wedge D(c)<D_{\mathrm{best}}-\epsilon
\label{eq:smt-opt}
\end{equation}
is satisfiable for a tolerance $\epsilon>0$.
A satisfying assignment replaces the incumbent,
and optimization continues until no configuration
improves the predicted distortion by more than
$\epsilon$.
The final incumbent is passed to live calibration
as the candidate configuration.

\subsection{Live Calibration}
\label{sec:measurement-backed}

The Performance Database characterizes the offline image
pool, which may differ from the user's images.
\sys{} therefore evaluates candidate configurations
through the complete encoder, attack, and decoder pipeline
on the user's images before deployment.

\paragraph{Live Check.}
For each user image $x_{o,i}$ and requested attack
$a\in\mathcal A_u$, the configured decoder $D_c$
evaluates the attacked watermarked image
$a(E_c(x_{o,i},w))$. Each readout is scored by its bit
accuracy $\operatorname{BA}$ against the expected
payload $w$.

The decoder first evaluates all individual-fragment
readouts and, when multiple fragments are selected,
the fused readout on the received image. If the highest
score reaches $\tau(c,\alpha)$, the image is accepted
with that score. Otherwise, if geometric recovery is
enabled, the decoder evaluates geometric readouts in
a fixed order and accepts at the first passing readout.
If no readout passes, the highest score from the received
image is retained. We denote the resulting verification
score by $s_i(c,a)$.

Given $N_a$ evaluated images and a required
image-acceptance rate $\rho\in(0,1]$, we require
\begin{equation}
\begin{aligned}
&\frac{1}{N_a}\sum_i
\mathbf{1}\big[s_i(c,a)\ge\tau(c,\alpha)\big]
\ge\rho,\\
&\frac{1}{N_a}\sum_i s_i(c,a)
\ge\tau(c,\alpha),
\qquad \forall a\in\mathcal A_u.
\end{aligned}
\label{eq:live-gate}
\end{equation}
These conditions enforce both image-level acceptance
and mean bit accuracy under every requested attack.
The measured embedding PSNR and deployment latency
must also satisfy the quality floor $q$ and latency
ceiling $t$, respectively.

\paragraph{Request-Local Refinement.}
Candidates are evaluated in ascending order of predicted
distortion, with up to three distinct discrete structures.
If none passes, the measured fragment accuracies and
image-level acceptance rates are used to apply
request-local downward corrections to the corresponding
performance predictions. The solver then proposes one
revised candidate, which must pass the same live check.
If no candidate passes under the current screening
margins, \sys{} retries with smaller margins according
to a predefined schedule. The deployment requirements,
FPR accounting, and live acceptance criteria remain
unchanged throughout refinement
(Appendix~\ref{sec:method-calibration-details}).

\paragraph{Outcome.}
The first candidate passing all live checks is returned
as $c_{\mathrm{out}}$, and \sys{} reports \textsc{sat}.
Deployment uses $E_{c_{\mathrm{out}}}$ to watermark
images and $D_{c_{\mathrm{out}}}$ to verify them at
threshold $\tau(c_{\mathrm{out}},\alpha)$.
If no candidate passes the live checks, \sys{}
reports \textsc{unsat}.

\section{Evaluation}
\label{sec:eval}
\begin{table*}[!t]
\centering
\caption{Request satisfaction and image fidelity across deployment scenarios.}
\label{tab:main}
\papertablestyle\footnotesize\renewcommand{\arraystretch}{1.0}
\begin{adjustbox}{max width=\linewidth}
\begin{tabular}{@{}l cccccc @{\hspace{6pt}} ccc@{}}
\toprule
\toprule
& \multicolumn{6}{c}{\textit{Request satisfaction (\%) $\uparrow$}} & \multicolumn{3}{c}{\textit{Image fidelity}} \\
\cmidrule(lr){2-7} \cmidrule(lr){8-10}
Method & S1 & S2 & S3 & S4 & S5 & \textbf{Avg.} & PSNR $\uparrow$ & SSIM $\uparrow$ & PSNR-P \\
\midrule
StegaStamp & 0.00 & 0.00 & 0.00 & 0.00 & 0.00 & \cellcolor{orange!8}0.00 & - & - & - \\
MaskWM & 10.32 & 0.00 & 0.00 & 0.00 & 0.00 & \cellcolor{orange!8}2.06 & 38.51 & 0.9774 & 38.51 / 47.03 \\
VINE & 8.83 & 0.00 & 42.68 & 54.26 & 0.00 & \cellcolor{orange!8}21.15 & 35.01 & \underline{0.9931} & 35.01 / 40.26 \\
TrustMark & \textbf{100.00} & 0.00 & 23.38 & 0.00 & 0.00 & \cellcolor{orange!8}24.68 & 40.91 & 0.9909 & 40.91 / 43.55 \\
VideoSeal & 27.98 & 49.74 & 4.01 & 0.00 & 2.74 & \cellcolor{orange!8}16.89 & \textbf{45.16} & \textbf{0.9949} & 45.16 / 45.96 \\
\midrule
VINE + Geo. & 8.83 & 0.00 & 42.68 & 62.65 & 0.00 & \cellcolor{orange!8}22.83 & 34.87 & 0.9921 & 34.87 / 39.92 \\
TrustMark + Geo. & \textbf{100.00} & 94.71 & 23.38 & 0.00 & 11.94 & \cellcolor{orange!8}46.01 & 40.08 & 0.9884 & 40.08 / 43.70 \\
\midrule
Optimized Single & \textbf{100.00} & \textbf{100.00} & 69.70 & 73.72 & 23.45 & \cellcolor{orange!8}73.37 & \underline{41.85} & 0.9915 & 41.85 / 42.36 \\
Always All Three & \underline{99.61} & \textbf{100.00} & \textbf{97.31} & \underline{77.74} & \underline{97.70} & \cellcolor{orange!8}\underline{94.47} & 38.98 & 0.9892 & 38.98 / 41.08 \\
\sysgreedy{} & \textbf{100.00} & \underline{97.39} & \underline{87.55} & 76.03 & 94.27 & \cellcolor{orange!8}91.05 & 40.91 & 0.9914 & 40.91 / 41.22 \\
\sysfull{} & \textbf{100.00} & \textbf{100.00} & \textbf{97.31} & \textbf{85.40} & \textbf{98.34} & \cellcolor{orange!8}\textbf{96.21} & 41.02 & 0.9923 & - \\
\bottomrule
\bottomrule
\end{tabular}
\end{adjustbox}
\papertablenote{PSNR-P: row method / \sysfull{}, evaluated on jointly accepted requests.}
\end{table*}

\subsection{Experiment Setup}
\label{sec:exp-setup}

\begin{figure*}[t]
\centering
\begin{minipage}[t]{0.325\linewidth}
\centering
\vspace{0pt}
\includegraphics[height=126pt]{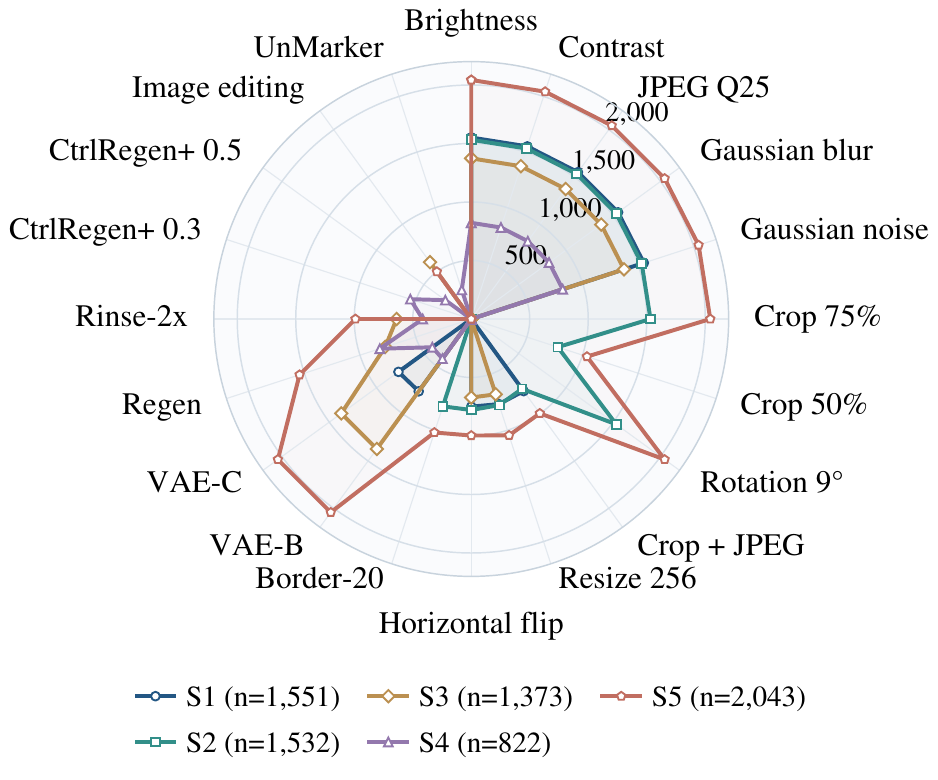}
\caption{Number of requests containing each attack across the five scenarios.}
\label{fig:attack-counts-by-scenario}
\end{minipage}\hfill
\begin{minipage}[t]{0.35\linewidth}
\centering
\vspace{0pt}
\includegraphics[height=126pt]{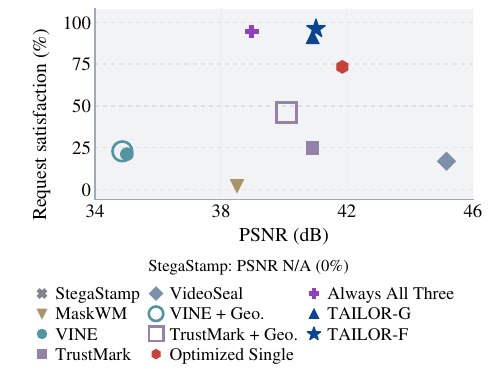}
\caption{Scenario-averaged request satisfaction and image fidelity.}
\label{fig:quality-overall}
\end{minipage}\hfill
\begin{minipage}[t]{0.31\linewidth}
\centering
\vspace{0pt}
\includegraphics[height=126pt]{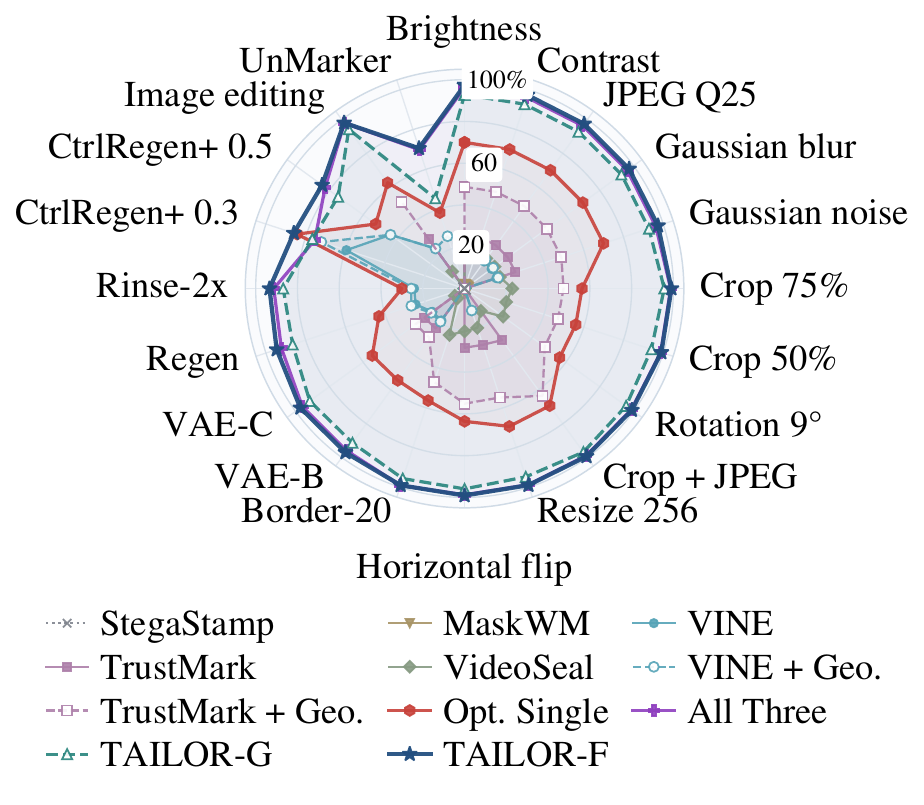}
\caption{Request satisfaction grouped by attack inclusion.}
\label{fig:request-radar-attacks}
\end{minipage}
\end{figure*}

\noindent\textbf{Datasets and Requests.}
\label{sec:datasets}
We construct an evaluation image pool from
MS-COCO~\citep{coco}, DiffusionDB~\citep{diffusiondb},
UltraEdit~\citep{ultraedit}, DIV2K~\citep{div2k},
and LSDIR~\citep{lsdir}, together with images
generated by DALL$\cdot$E~3 and SDXL.
These sources span natural photographs,
diffusion-generated images, and editing-oriented
content across diverse image characteristics.
Each deployment request specifies an attack set,
an FPR budget, an image-quality floor, and a
latency ceiling.
The benchmark comprises 7,321 distinct requests
covering 20 attack settings across five scenarios:
signal processing and re-encoding (S1),
geometric transformations (S2),
regeneration and editing (S3),
watermark removal (S4), and
broad attack coverage (S5).
\autoref{fig:attack-counts-by-scenario} shows the
distribution of requested attacks across scenarios,
where each request may include multiple attacks.
Scenario construction, image sampling,
deduplication, and attack settings are detailed
in Appendices~\ref{sec:eval-datasets-details}
and~\ref{sec:deployment-eval-details}.

\noindent\textbf{Baselines.}
\label{sec:baselines} \textbf{TAILOR-F} denotes the full version of \sys{}.
We compare it with five standalone watermark methods:
\textbf{MaskWM}~\citep{maskwm},
\textbf{StegaStamp}~\citep{stegastamp},
\textbf{VINE}~\citep{vine},
\textbf{TrustMark}~\citep{trustmark}, and
\textbf{VideoSeal}~\citep{videoseal},
all at unit embedding strength, to assess the
limitations of individual watermarks.
Geometric variants (Geo.) of VINE and TrustMark
further evaluate the benefit of geometric recovery.
We also include three optimized baselines.
\textbf{Optimized Single} selects one fragment per
request and optimizes its strength and geometric
recovery, testing the need for composition.
\textbf{Always All Three} fixes all three fragments
while optimizing their strengths, order, and geometric
recovery, testing the benefit of adaptive fragment
selection.
\textbf{\sysgreedy{}} incrementally selects fragments
by predicted coverage gain without backtracking,
testing the benefit of joint configuration search.
These baselines share the same performance database,
parameter optimization, and live validation as
TAILOR-F, subject to their selection policies.
All methods use identical requests, images, and
acceptance criteria. Further details are provided
in Appendix~\ref{sec:eval-baseline-details}.

\noindent\textbf{Evaluation Metrics.}
\label{sec:metrics}
We evaluate \sys{} using \textbf{request satisfaction}
and \textbf{image fidelity}.
\textbf{Request satisfaction} measures the percentage
of deployment requests that pass live validation
under every requested attack while satisfying the
FPR budget, image-quality floor, and latency ceiling.
We report satisfaction rates for each scenario and
their unweighted average.
\textbf{Image fidelity} is measured by
\textbf{PSNR}~\citep{huynh2008scope} and
\textbf{SSIM}~\citep{wang2004ssim} between original
and watermarked images before attack, over each
method's satisfied requests.
PSNR measures pixel-level distortion, while SSIM
measures structural similarity.
Since methods may satisfy different request sets,
we additionally report \textbf{PSNR-P}, which compares
the mean PSNR of \sys{} and each baseline on their
jointly satisfied requests, enabling fidelity
comparisons under matched deployment requirements.

\subsection{Experimental Results}
\label{sec:exp-results}

\noindent\textbf{Request Satisfaction.}
We evaluate whether \sys{} satisfies complete deployment
requests across different attack scenarios.
As shown in~\autoref{tab:main}, \sys{} achieves a
scenario-averaged satisfaction rate of 96.21\%,
outperforming Optimized Single (73.37\%),
Always All Three (94.47\%), and \sysgreedy{}
(91.05\%) by 22.84, 1.74, and 5.16 percentage
points, respectively.
Optimized Single satisfies all S1 and S2 requests
but only 23.45\% of S5 requests, compared with
98.34\% for \sys{}, demonstrating the importance
of combining complementary watermarks.
Always All Three achieves strong coverage but
falls to 77.74\% on S4, compared with 85.40\%
for \sys{}, highlighting the benefit of adaptive
fragment selection.
The improvement over \sysgreedy{} further
demonstrates the value of joint configuration
optimization.
\Cref{fig:request-radar-attacks} illustrates
\sys{}'s consistently strong coverage across
individual attack groups.

\noindent\textbf{Image Fidelity.}
As shown in~\autoref{tab:main}, \sys{} achieves
a mean PSNR of 41.02~dB and an SSIM of 0.9923.
\autoref{fig:quality-overall} summarizes its
request satisfaction and image fidelity.
Some standalone methods achieve higher mean PSNR,
such as VideoSeal at 45.16~dB, but satisfy
substantially fewer requests.
We therefore use \textbf{PSNR-P} to compare
fidelity on jointly satisfied requests.
Under this paired comparison, \sys{} achieves
42.36~dB versus 41.85~dB for Optimized Single,
41.08~dB versus 38.98~dB for Always All Three,
and 41.22~dB versus 40.91~dB for \sysgreedy{}.
These gains of 0.51, 2.10, and 0.31~dB,
respectively, show that joint optimization
improves image fidelity while maintaining broader
request coverage.
In particular, the improvement over Always All
Three highlights the quality benefit of adapting
the watermark composition to each request.

\noindent\textbf{Watermark Selection.}
We select VINE, TrustMark, and VideoSeal for their
complementary robustness across attack scenarios.
Among fixed-strength standalone watermarks without
geometric recovery, TrustMark achieves the highest
satisfaction on S1 (100\%), VideoSeal on S2
(49.74\%), and VINE on S3 and S4 (42.68\% and
54.26\%). These complementary strengths motivate
their joint composition for heterogeneous attack
requirements.
We further assess their compatibility through
direct superposition
(Appendix~\ref{sec:fragment-superposition}).
Under unit-strength superposition without attacks,
mean raw bit accuracy decreases by only 0.49
percentage points across the tested compositions.
The results also reveal order-dependent interference,
particularly for VideoSeal, motivating the joint
optimization of fragment selection and embedding
order. Further analysis of recovery interactions
is provided in
Appendix~\ref{sec:recovery-interaction-ablation}.

\begin{table}[t]
\centering
\caption{Ablation of candidate search and strength adaptation.}
\label{tab:ablation-process}
\papertablestyle\footnotesize\renewcommand{\arraystretch}{1.0}
\begin{adjustbox}{max width=\linewidth}
\begin{tabular}{@{}l c ccc@{}}
\toprule
\toprule
& \textit{Satisfaction} & \multicolumn{3}{c}{\textit{Image fidelity}} \\
\cmidrule(lr){2-2} \cmidrule(lr){3-5}
Method & Rate (\%) $\uparrow$ & PSNR $\uparrow$ & SSIM $\uparrow$ & PSNR-P \\
\midrule
Unit-strength & 73.18 & 37.85 & 0.9887 & 37.92 / 41.40 \\
SMT-top1 & 80.65 & \underline{41.22} & \textbf{0.9927} & 41.22 / 41.22 \\
SMT-top3 & \underline{91.99} & 40.83 & \underline{0.9920} & 40.83 / 40.83 \\
\sysfull{} & \textbf{95.50} & 40.66 & 0.9918 & - \\
\bottomrule
\bottomrule
\end{tabular}
\end{adjustbox}
\papertablenote{PSNR-P: row variant / \sysfull{}, evaluated on jointly accepted requests.}

\end{table}

\subsection{Ablation Studies}
\label{sec:ablation}

\noindent\textbf{Candidate Search and Live Refinement.}
We evaluate whether broader candidate search and live feedback improve request
satisfaction. As shown in Table~\ref{tab:ablation-process}, SMT-top1 satisfies 80.65\% of requests, while evaluating up to
three candidate structures increases satisfaction to 91.99\%, a gain of 11.34
percentage points. Request-local refinement and margin fallback further raise
satisfaction to 95.50\%, adding another 3.51 points. This shows that the
highest-ranked offline candidate is not always sufficient and that broader
search and live refinement recover additional feasible deployments.
On jointly accepted requests, SMT-top1 and SMT-top3 select the same
configurations as \sys{}, yielding identical mean PSNRs of 41.22 and
40.83~dB on their respective intersections. Thus, the additional search expands
coverage without degrading fidelity on already successful requests.

\noindent\textbf{Embedding Strength Optimization.}
We evaluate the benefit of adaptive embedding strengths
by first fixing all selected strengths to one while
retaining the other configuration decisions and live
refinement. Unit-strength satisfies 73.18\% of requests,
compared with 95.50\% for \sys{}.
On jointly satisfied requests, \sys{} improves PSNR
from 37.92 to 41.40~dB (+3.48~dB), demonstrating
the importance of per-request strength adaptation.
We further compare continuous optimization with
Grid, which restricts strengths to measured knots
(Appendix~\ref{sec:grid-enumeration}).
Continuous optimization improves paired PSNR by
0.33~dB on 7,130 jointly satisfied requests.
For the frozen deployments, PSNR-threshold analysis
retains 3.9 and 7.7 percentage points more requests
at 40 and 42.6~dB, respectively
(Figure~\ref{fig:fidelity-sweep}).
Together, these results demonstrate the benefits of
both adapting embedding strengths to deployment
requirements and optimizing them continuously.
Additional ablations are in
Appendix~\ref{sec:eval-more-results}.

\section{Limitations}
\sys{} currently evaluates three pretrained watermark
fragments under 20 attack settings. Extending the
fragment library requires compatible payload and
verification interfaces, together with additional
characterization of recovery, distortion, latency, and
composition interactions.

The optimization relies on response models derived from
offline measurements, whose prediction errors may affect
configuration selection. Live calibration mitigates
these errors by validating candidate configurations on
the deployment images and refining inaccurate predictions.

Our evaluation covers a broad range of existing
watermark attacks, including signal processing,
geometric transformations, neural compression,
regeneration, image editing, and optimization-based
watermark removal. While these attacks represent
diverse and practically relevant threats, the
generalization of \sys{} to previously unseen
attack methods remains an important direction
for future investigation.

\section{Conclusion}
\label{sec:conclusion}
We present \sys{}, a request-conditioned framework that jointly
optimizes watermark selection, embedding strengths, order, and
geometric recovery under deployment-specific robustness, FPR,
image-quality, and latency requirements. By combining offline
characterization, SMT-based optimization, and live calibration,
\sys{} adapts watermark compositions to diverse deployment needs.
Experiments on 7,321 requests across five scenarios and 20 attack
settings achieve 96.21\% scenario-averaged request satisfaction
with a mean PSNR of 41.02~dB. Comparisons with optimized single-
and three-watermark baselines demonstrate that adaptive composition
improves attack coverage while preserving image fidelity.
These results highlight the effectiveness of jointly optimizing
complementary watermarks to meet practical deployment requirements.

{
    \small
    \bibliographystyle{ieeenat_fullname}
    \bibliography{main}
}

\RemoveFromHook{para/begin}[tailor-fit]

\clearpage
\fitoff
\showrevfalse
\setcounter{page}{1}
\maketitlesupplementary
\appendix

\section{Implementation Details}
\label{sec:method-details}

This appendix provides the implementation details for the three stages of
\sys{}, including the performance database, SMT encoding, FPR accounting,
live calibration, and watermark deployment. 
For reference, Table~\ref{tab:notation} summarizes the notation used throughout
the formulation and the main text.

\subsection{Preliminaries and Notation}
\label{sec:method-preliminaries}

\paragraph{Piecewise-Linear Response Curves.}
Strength-dependent recovery and distortion measurements
are represented using piecewise-linear interpolation
between measured knots (Eq.~\ref{eq:pwl}). Runtime is modeled
separately using measured fragment and geometric-stage
costs (Appendix ~\ref{sec:method-table-details}).For consecutive knots
$\lambda^{(0)}<\cdots<\lambda^{(K)}$, the SMT encoding introduces one Boolean
variable $z_k$ for each interval and enforces exactly one active segment:
\begin{equation}
\begin{aligned}
&\bigvee_{k=0}^{K-1} z_k,
\qquad
\neg(z_k\wedge z_{k'}) \quad \forall\, k\neq k',\\
&z_k\Rightarrow
\Big[
\lambda^{(k)}\le\lambda\le\lambda^{(k+1)}\\
&\qquad\quad\wedge\
\widehat y
=
y^{(k)}+s^{(k)}\big(\lambda-\lambda^{(k)}\big)
\Big].
\end{aligned}
\label{eq:pwl-encoding}
\end{equation}
The knots $\lambda^{(k)}$, measured values $y^{(k)}$, and slopes $s^{(k)}$
are fixed constants, so every active segment contributes only linear
constraints over the strength variable. A separate constraint confines each
strength to the characterized range of its fragment
(Appendix~\ref{sec:native-strength}), so the solver interpolates between
measured strengths and never selects a strength outside that range. A few
curves are profiled over a narrower span than this range, namely the pairwise
distortion correction, the scale-stage distortion, and VINE recovery under
UnMarker. The encoding holds such a curve constant at its end value beyond its
outermost knot.

Recovery curves $b_f(\cdot,a)$ and interaction curves
$\delta_{g\to f}(\cdot,a)$ are maintained separately for each attack.
Standalone distortion $D_f(\cdot)$ and latency are modeled per fragment, while
the pairwise distortion correction $e_{fg}$ is indexed by
$\lambda_f+\lambda_g$.

\paragraph{SMT Formulation.}
Configuration selection couples discrete structural choices with continuous
embedding strengths. Fragment selection, embedding order, and geometric
recovery are discrete decisions, whereas each selected fragment has a
continuous native strength; these variables interact because predicted
recovery and distortion depend on the complete composition. SMT~\citep{barrett2018satisfiability} is therefore
well suited to the problem, since it combines Boolean structure with arithmetic
constraints in a single model.

We encode all request requirements and configuration-validity conditions as an
SMT formula over Boolean variables and linear real arithmetic. A satisfying
assignment directly specifies a feasible configuration, while a solver-level
\textsc{unsat} result indicates that no configuration in the characterized
search space satisfies the encoded requirements at the current screening
margins. This differs from the pipeline-level \textsc{unsat} that \sys{}
reports for a request (Sec.~\ref{sec:measurement-backed}), which means that no
candidate passed live validation within the candidate and fallback budgets of
Appendix~\ref{sec:method-calibration-details}. We use Z3~\citep{z3} to
solve the resulting model.

Optimization is performed by repeated satisfiability queries. After obtaining a
feasible configuration with objective value $D_{\mathrm{best}}$, we add the
constraint that the next solution must improve distortion by at least
$\epsilon$ and solve again, with $\epsilon=10^{-9}$ in MSE. This process
continues until no improving assignment exists, yielding the lowest-distortion
configuration up to the specified tolerance.

\begin{table*}[!htbp]
\centering
\caption{Core notation used throughout the method and appendix.}
\label{tab:notation}
\papertablestyle
\begin{tabular}{@{}>{$}l<{$} p{\dimexpr\linewidth-3.1cm-2\tabcolsep\relax}@{}}
\toprule
\textrm{Symbol} & Meaning \\
\midrule

\multicolumn{2}{@{}l}{\textit{Deployment request}} \\
u=(\mathcal A_u,\alpha,q,t)
& Deployment request specifying the attack set, FPR budget, PSNR floor, and latency ceiling \\
\mathcal A_u
& Set of requested attacks; $a$ denotes one attack \\
\alpha
& Configuration-level false-positive-rate budget for one image verification \\
q,\ t
& Minimum PSNR in dB and maximum deployment latency in ms \\

\midrule
\multicolumn{2}{@{}l}{\textit{Configuration}} \\
f,\ g
& Watermark fragments drawn from VINE, TrustMark, and VideoSeal \\
c=(S,\lambda,\pi,\varphi)
& Configuration specifying the selected fragments, native strengths, embedding order, and geometric recovery stage \\
S,\ m=|S|
& Selected fragment subset and number of selected fragments \\
\lambda_f
& Native embedding strength of fragment $f$ within its characterized range \\
\pi,\ f\prec_\pi g
& Embedding order; $f\prec_\pi g$ means that $g$ is embedded after $f$ \\
\varphi
& Enabled geometric recovery stage, at most one of scale, tile, resync, and angle \\
E_c,\ D_c
& Composite encoder and configured decoder induced by $c$ \\

\midrule
\multicolumn{2}{@{}l}{\textit{Performance database}} \\
b_f(\lambda_f,a)
& Measured standalone mean bit accuracy of fragment $f$ at strength $\lambda_f$ under attack $a$ \\
\delta_{g\to f}(\lambda_g,a)
& Measured recovery loss of $f$ caused by embedding $g$ after it \\
\widehat b_f(c,a)
& Predicted mean bit accuracy of fragment $f$ under configuration $c$ and attack $a$ \\
\Gamma_{f,a}(c)
& Whether $f$ is predicted to satisfy the screening acceptance-rate requirement $\bar\rho$ under attack $a$ \\
D_f(\lambda_f)
& Measured standalone embedding MSE of fragment $f$ \\
e_{fg}(\lambda_f+\lambda_g)
& Measured pairwise excess MSE beyond the sum of standalone distortions \\
D_\varphi(c)
& Additional embedding MSE introduced by the enabled geometric recovery stage \\
D(c),\ P(c)
& Predicted total embedding MSE and corresponding PSNR \\
L(c)
& Predicted deployment latency of configuration $c$ in ms \\

\midrule
\multicolumn{2}{@{}l}{\textit{Verification}} \\
w,\ \ell_j
& Expected 100-bit codeword and aligned decoder logit for bit $j$ \\
\rev{\operatorname{BA}(w,\ell)}
& Bit-match accuracy of one aligned readout against $w$ \\
T(c)
& Maximum number of verification tests evaluated by configuration $c$ \\
\theta(c,\alpha),\ \tau(c,\alpha)
& Integer match threshold and its normalized value $\theta/100$; $\theta=101$ denotes an impossible-pass threshold \\
\rho
& Required fraction of images that must pass verification under each requested attack \\
\eta,\ \eta_{\mathrm{rate}}
& Screening margins applied to the recovery threshold and acceptance rate during selection \\

\midrule
\multicolumn{2}{@{}l}{\textit{Live calibration}} \\
s_i(c,a),\ N_a
& Verification score for image $i$ under configuration $c$ and attack $a$, and the number of evaluated images \\
\bar b,\ \mathrm{se}
& Live mean recovery and its standard error \\
\Delta b
& Request-local downward correction applied to an optimistic recovery prediction \\
c_{\mathrm{out}}
& Configuration accepted after live calibration for deployment \\

\bottomrule
\end{tabular}
\end{table*}

\subsection{Performance Database}
\label{sec:method-table-details}

This section gives the implementation details of the offline quantities
introduced in Sec.~\ref{sec:method-offline}, including how recovery
interactions, distortion corrections, geometric responses, and runtime are
represented in the Performance Database. Each strength-dependent quantity is
stored as the PWL response described in
Sec.~\ref{sec:method-preliminaries}, together with the per-image measurements
needed by request-local calibration.

\paragraph{Recovery.}
For each fragment-attack pair, the database stores the standalone response
$b_f(\lambda_f,a)$ and the corresponding per-image scores. Ordered
composition effects are represented by
$\delta_{g\to f}(\lambda_g,a)$, the loss in $f$'s recovery when
$g$ is embedded after it. To measure it, we embed $f$ at one fixed mid-range
strength, 0.6 for VINE and 1.0 for TrustMark and VideoSeal, embed $g$ after it
at each of $g$'s knots, and apply attack $a$. The loss is the mean bit
accuracy of $f$ alone minus its mean bit accuracy in the pair, and negative
values, which would be gains, are set to zero. The curve thus depends on
$\lambda_g$ only and is applied at every $\lambda_f$. The interaction under
UnMarker was not profiled and is set to zero. For a configuration with order
$\pi$, Eq.~\ref{eq:recovery} subtracts the measured correction for every
$g$ satisfying $f\prec_\pi g$. Supported geometric recovery stages maintain
separate response curves for the affected fragment-attack pairs. For some of
these pairs, the database also holds the stage response measured with a second
fragment embedded afterwards, which then replaces the single-fragment stage
response together with the corresponding $\delta$ term.
\paragraph{Image Quality.}
The database stores standalone embedding distortion $D_f(\lambda_f)$, the mean
over the characterization images of the per-image MSE on the 0 to 255 scale,
and a pairwise correction $e_{fg}$ for deviations from additive distortion
under sequential embedding. To keep the correction univariate, $e_{fg}$ is
indexed by $\lambda_f+\lambda_g$. It is measured on a $3\times3$ grid of a low,
a middle, and a high strength of each fragment (0.2, 0.6, and 1.0 for VINE;
0.4, 1.0, and 1.6 for TrustMark; 0.5, 1.0, and 1.5 for VideoSeal), with the
pair embedded in one fixed order, as the MSE of the pair minus the two
standalone MSEs. The nine strength pairs have nine distinct sums, which index
the correction, and negative values are set to zero. The
geometric-stage term $D_\varphi(c)$ is an added MSE. Scale adds a curve in
$\lambda_{\mathrm{VINE}}$ for its two additional VINE copies, from 1.2 MSE at
$\lambda_{\mathrm{VINE}}=0.2$ to 14.6 MSE at 1.0. Resync adds about 2.6 MSE
for the synchronization template. Tile and angle add nothing: the tiled
TrustMark embedding measured less distortion than the plain one, a
difference that the model sets to zero, and angle does not change the
embedding. Together, these quantities instantiate Eq.~\ref{eq:distortion}.

The distortion model is independent of the embedding order: neither $D_f$ nor
$e_{fg}$ carries an order index, and there is no three-fragment term. Order
enters the selection through the recovery model, which determines the
strengths that a given order requires. The pairwise correction is small for
the deployed library. For TrustMark-VINE, the correction decreases from
0.35 MSE at the smallest strength sum to 0.10 MSE at the largest. For
TrustMark-VideoSeal and VINE-VideoSeal, the measured excess is slightly
negative at all nine strength pairs, between $-0.26$ and $-0.02$ MSE, and the
model sets it to zero. For comparison, standalone distortion ranges from about
0.6 to 21 MSE over the characterized strengths.

\begin{revblock}
\paragraph{Distortion Model Accuracy.}
\label{sec:distortion-accuracy}
We compare selection-time PSNR predictions from the offline Performance
Database with mean embedding PSNR measured on the 100 live images, which are
disjoint from the characterization slices
(Appendix~\ref{sec:eval-datasets-details}). Each sample is a distinct deployed
configuration, identified by its fragments, order, geometric stage, and exact
strengths. De-duplication is global across methods: a configuration deployed
by \sysfull{} and by a control is counted once, under \sysfull{}, so the
1,704 control configurations of Table~\ref{tab:distortion-accuracy} are those
that only a control deploys. Across the 2,393
configurations deployed by \sysfull{} or the two
composition controls, the mean absolute error is 0.095~dB, the 95th percentile
is 0.224~dB, and the maximum is 0.511~dB; the signed mean is $+0.027$~dB
(Table~\ref{tab:distortion-accuracy}). Mean absolute error is 0.186~dB with
geometry and 0.084~dB without it; geometric configurations account for 8 of the
10 errors above 0.3~dB. Scale and tile stages exhibit signed mean errors of
$+0.23$ and $-0.30$~dB, respectively. Live calibration checks each candidate's
measured PSNR against the requested floor before deployment.
\end{revblock}

\begin{table}[!htbp]
\centering
\revcolor
\caption{Prediction error of the embedding PSNR of deployed configurations, in dB.}
\label{tab:distortion-accuracy}
\papertablestyle
\begin{adjustbox}{max width=\linewidth}
\begin{tabular}{@{}lrrrrrr@{}}
\toprule
& & & \multicolumn{3}{c}{\textit{Absolute error}} & \\
\cmidrule(lr){4-6}
Configurations & $n$ & Signed mean & Mean & 95th pct. & Max & $>0.3$ \\
\midrule
All deployed & 2,393 & $+0.027$ & 0.095 & 0.224 & 0.511 & 10 \\
\quad \sysfull{} & 689 & $-0.059$ & 0.105 & 0.236 & 0.511 & 6 \\
\quad composition controls & 1,704 & $+0.062$ & 0.090 & 0.200 & 0.510 & 4 \\
\midrule
One fragment & 138 & $-0.103$ & 0.149 & 0.493 & 0.511 & 9 \\
Two fragments & 508 & $-0.070$ & 0.091 & 0.196 & 0.248 & 0 \\
Three fragments & 1,747 & $+0.066$ & 0.091 & 0.234 & 0.306 & 1 \\
\midrule
No geometric stage & 2,145 & $+0.017$ & 0.084 & 0.169 & 0.396 & 2 \\
With a geometric stage & 248 & $+0.121$ & 0.186 & 0.292 & 0.511 & 8 \\
\quad scale & 149 & $+0.232$ & 0.232 & 0.291 & 0.306 & 1 \\
\quad tile & 22 & $-0.303$ & 0.303 & 0.510 & 0.511 & 7 \\
\quad resync & 20 & $-0.013$ & 0.057 & 0.135 & 0.135 & 0 \\
\quad angle & 57 & $+0.043$ & 0.065 & 0.136 & 0.169 & 0 \\
\bottomrule
\end{tabular}
\end{adjustbox}
\papertablenote{Error: predicted minus measured PSNR. The last column counts absolute errors above 0.3~dB.}
\end{table}

\paragraph{Runtime.}
The latency model sums the measured embedding and decoding costs of the
selected fragments and adds the overhead of the enabled geometric recovery
stage. The per-fragment costs are 57.4~ms for VINE, 12.4~ms for TrustMark, and
22.4~ms for VideoSeal. The stage overhead is the additional decoding time
measured offline with the stage enabled. It is counted once per configuration,
at its largest value over the requested attacks, because the stage search runs
at most once per image. These offline measurements instantiate the predicted
latency $L(c)$ that the solver uses during selection. The measured deployment
latency is defined in Appendix~\ref{sec:eval-metric-details}.

\subsection{Decoder and FPR Accounting}
\label{sec:method-budget-details}

A configuration may expose multiple verification paths through individual
fragments, fused readouts, and geometric recovery views. Evaluating more paths
can improve watermark recovery but also increases the opportunity for false
positives. \sys{} therefore uses a common verification score for all readouts
and accounts for every permitted test when setting the detection threshold.

\paragraph{Unified Verification Score.}
Let $w\in\{0,1\}^{100}$ denote the expected watermark codeword and $\ell_j$ the
aligned decoder logit for bit position $j$. TrustMark and VideoSeal return
real-valued logits, and VINE returns sigmoid probabilities $p$, which we
convert to $\ln(p/(1-p))$. No further scaling or calibration is applied, so
these values are decoder logits rather than calibrated log-likelihood ratios.
Every readout is scored by its bit accuracy against $w$:
\begin{equation}
\rev{\operatorname{BA}(w,\ell)}
=
\frac{1}{100}
\sum_{j=1}^{100}
\mathbf{1}\!\left[(2w_j-1)\ell_j>0\right].
\label{eq:unified-score}
\end{equation}
Each fragment's key-derived permutation and sign mask are inverted before
scoring. When multiple fragments are selected, the decoder also forms one
fused readout on the received image, the sum of their aligned logits, and
scores it with the same rule. Individual,
fused, and geometric readouts therefore share a common verification statistic.

\begin{revblock}
\paragraph{Readouts.}
Verification under identity $v$ reads three kinds of aligned readouts, which we
write in the notation of Appendix~\ref{sec:fpr-selection}. An \emph{individual
readout} takes the raw logits $y_f\in\mathbb R^{100}$ that the
decoder of one selected fragment $f$ returns on one view and aligns them with
that fragment's key-derived permutation $\sigma_{f,v}$ and sign mask
$M_{f,v}$, giving $\ell_{f,j}=M_{f,v}[\sigma_{f,v}(j)]\,y_f[\sigma_{f,v}(j)]$
(Eq.~\ref{eq:selection-aligned-llr}). The \emph{fused readout} exists only on
the primary view, the image as received, and only when $m>1$. It adds the
aligned logits of all selected fragments,
$\ell_{\mathrm{fuse},j}=\sum_{f\in S}\ell_{f,j}$. \emph{Geometric readouts}
are individual readouts on the candidate views of the enabled stage
(Appendix~\ref{sec:geometric-stages}): 133 centred crops read by VINE for
scale, 81 windows read by TrustMark for tile, the rectified frame and four
tilts read by every selected fragment for resync, and at most 65 rotations read
by every selected fragment for angle. The candidate views are computed from the
image alone. The angle candidates, for example, come from the corner wedges of
the image followed by a fixed sweep. The views depend neither on the key nor on
any match count, and geometric views are never fused. Every readout $r$ is
scored by its bit accuracy $\operatorname{BA}_r=\operatorname{BA}(w,\ell_r)$,
equivalently by its match count $K_r=100\,\operatorname{BA}_r$.

\paragraph{Accept Decision and Candidate Tests.}
An image is accepted if and only if some executed readout reaches the
threshold, $K_r\ge\theta(c,\alpha)$, with $\theta$ from
Eq.~\ref{eq:unified-threshold}. The decoder first scores all individual
readouts of the primary view and, when $m>1$, their fusion. If the highest of
these scores reaches the threshold, the image is accepted with that score.
Only otherwise does it walk the candidate views of the enabled geometric stage
in a fixed order, and it stops at the first readout that passes. Early stopping
decides which tests are executed but never adds a test. Let
$\mathcal R(c,x)$ denote the candidate tests of configuration $c$ on image
$x$: the primary individual readouts, the primary fusion when $m>1$, and every
geometric readout listed above. This set is fixed by $(c,x)$ before any match
count is computed, the executed tests form a subset of it, and
$|\mathcal R(c,x)|\le T(c)$ for the budget defined next.
\end{revblock}

\paragraph{Verification-Test Budget.}
Let $T(c)$ denote the maximum number of verification tests permitted by
configuration $c$:
\begin{equation}
\begin{aligned}
T(c)
&=
T_0(m)
+5m z_r
+133 z_s
+81 z_d
+65m z_a,\\
T_0(m)
&=
\begin{cases}
1, & m=1,\\
m+1, & m>1,
\end{cases}
\end{aligned}
\label{eq:unified-budget}
\end{equation}
where $m=|S|$, and $z_r,z_s,z_d,z_a\in\{0,1\}$ indicate whether
resynchronization, scale search, tile search, and angle search are enabled,
respectively. The base term $T_0(m)$ counts the individual fragment readouts
and, when $m>1$, one additional fused readout. The remaining terms count the
candidate views introduced by the corresponding geometric recovery stages.
Thus, the complete verification-test budget is determined by the selected
configuration before an image is evaluated. \rev{For example, VINE and
VideoSeal with scale search have $T(c)=3+133=136$ tests, and all three
fragments with resynchronization have $T(c)=4+15=19$.}

\paragraph{Null Model.}
We calibrate the verification threshold under the null hypothesis that the
evaluated image does not contain the deployed watermark. Let $\Pr_0$ denote
probability under this null, averaging over secret-key generation and negative
images for a fixed configuration and watermark codeword.

For each permitted readout $r$, let $J_r$ denote the bit positions with nonzero
aligned logits. We assume that, under the null, the signs at these
positions are mutually independent and equally likely to agree or disagree
with the expected codeword. If $K_r=100\,\rev{\operatorname{BA}_r}$ denotes the number of matching
bits, then
\begin{equation}
\begin{aligned}
&K_r\mid J_r\sim\mathrm{Bin}(|J_r|,1/2),\\
&\Pr_0[K_r\ge h]
\le
\Pr[\rev{B}\ge h],
\quad
\rev{B}\sim\mathrm{Bin}(100,1/2).
\end{aligned}
\label{eq:null-binomial}
\end{equation}
Zero logits are counted as mismatches, so the 100-bit binomial tail
remains an upper bound when some bit positions are unresolved. This assumption
concerns the aligned decoder outputs with the expected codeword held fixed and
does not require independence among the bits of the codeword itself.

\paragraph{Configuration-Level FPR Control.}
Because configuration $c$ may evaluate up to $T(c)$ verification tests, we
allocate each test a false-positive budget of $\alpha/T(c)$. We then choose the
smallest number of matching bits whose binomial upper-tail probability satisfies
this budget:
\begin{equation}
\begin{aligned}
\theta(c,\alpha)
&=
\min\left\{
h\in\{0,\ldots,101\}:
\Pr[\rev{B}\ge h]\le\frac{\alpha}{T(c)}
\right\},\\
\tau(c,\alpha)&=\frac{\theta(c,\alpha)}{100}.
\end{aligned}
\label{eq:unified-threshold}
\end{equation}
The sentinel value $h=101$ represents an impossible-pass threshold when no
attainable 100-bit match count satisfies the requested per-test budget. A
verification test passes only when its score reaches $\tau(c,\alpha)$.

Under the null model, every permitted test therefore has false-positive
probability at most $\alpha/T(c)$. Applying a union bound over all permitted
tests gives
\begin{equation}
\Pr_0(\mathrm{accept})
\le
\sum_{r=1}^{T(c)}
\Pr_0\!\left[\rev{\operatorname{BA}_r}\ge\tau(c,\alpha)\right]
\le
T(c)\frac{\alpha}{T(c)}
=
\alpha.
\label{eq:fpr-accounting}
\end{equation}
\rev{A configuration with more verification paths receives a threshold that is
at least as high, which maintains the requested configuration-level FPR budget
under the null model.}

\begin{revblock}
\paragraph{Interpretation.}
Equation~\ref{eq:fpr-accounting} is a model-based FPR bound, rather than an
empirical estimate. Deployment uses one frozen configuration
$c_{\mathrm{out}}$; under the assumptions of
Proposition~\ref{prop:fpr-selection} in Appendix~\ref{sec:fpr-selection},
calibration retries do not add verification tests to $T(c_{\mathrm{out}})$.
\end{revblock}

\subsection{Per-Image Acceptance}
\label{sec:method-acceptance-details}

Offline selection uses both predicted mean recovery and stored image-level
acceptance rates. The indicator $\Gamma_{f,a}(c)$ is computed from the stored
per-image bit accuracies of fragment $f$ embedded alone under attack $a$. At
each knot, the acceptance rate is the fraction of these scores reaching the
threshold $\tau(c,\alpha)$ of the candidate configuration, which depends on
$c$ only through the number of selected fragments and the enabled stage
(Eq.~\ref{eq:unified-budget}). A nondecreasing isotonic fit over the knots,
interpolated linearly, gives the smallest strength $\lambda^{*}$ at which the
fitted rate reaches $\bar\rho=\min(1,\rho+\eta_{\mathrm{rate}})$, and
$\Gamma_{f,a}(c)$ holds if and only if $\lambda_f\ge\lambda^{*}$. If the
fitted rate never reaches $\bar\rho$, fragment $f$ cannot cover attack $a$.
The composition corrections $\delta$ do not enter $\Gamma$. They enter the
mean condition $\widehat b_f(c,a)\ge\bar\tau(c,\alpha)$, which
Eq.~\ref{eq:cov} imposes jointly with $\Gamma$. When the enabled stage changes
how $f$ is embedded or read under $a$, the per-image scores measured with
that stage are used instead. We use $\rho=0.90$, initial rate margins
$\eta_{\mathrm{rate}}\in\{0.04,0.06\}$, and initial mean-accuracy margin
$\eta=0.02$. Clean-image sweeps additionally impose minimum strengths
(Appendix~\ref{sec:native-strength}).

\subsection{Request-Local Refinement}
\label{sec:method-calibration-details}

Offline predictions may not transfer accurately to the user's images. Live
calibration therefore uses failed candidate evaluations to identify where the
performance database is optimistic, applies conservative request-specific
corrections, and re-solves the configuration problem. The correction is local
to the current request and affects only the fragment-attack cells that the
failed candidates measured.

\paragraph{Live Discrepancy Measurement.}
At each margin level, the solver returns up to three candidates with distinct
discrete structures, in ascending order of predicted distortion and each with
optimized continuous strengths. A structure consists of the selected fragment subset,
embedding order, and geometric recovery stage.
Different embedding orders of the same subset are
treated as distinct structures. The candidates are validated in this
order, and the first one that passes is deployed. Live measurements are cached
per configuration and attack and are reused across requests and margin levels.
If all candidates fail, each selected fragment $f$ and requested attack $a$ of
a failed candidate provides the per-image bit accuracies of $f$'s own readout
at the strength $\lambda_f$ used by that candidate. They are read on the
received image, or on the geometric view that accepted the image when a
geometric readout did. We compare two live statistics of fragment $f$ with the
corresponding offline predictions: its mean bit accuracy $\bar b$, with
standard error $\mathrm{se}=\hat\sigma/\sqrt{N_a}$, and its image-level
acceptance rate, i.e., the fraction of images on which $f$'s own readout
reaches the verification threshold $\tau(c,\alpha)$. The configuration-level
score $s_i(c,a)$ of Eq.~\ref{eq:live-gate}, which may come from the fused
readout or from a geometric view, decides the live check only and is not used
for these corrections.

\paragraph{Mean-Recovery Correction.}
Let $\widehat b$ be the recovery of fragment $f$ predicted by the database at
the same configuration and strength (Eq.~\ref{eq:recovery}), and define the
observed gap
$g=\bar b-\widehat b$. We do not directly replace the offline prediction with
the live mean, since the live measurement itself is estimated from a finite
image set. Instead, we apply the shrinkage correction
\begin{equation}
\Delta b=
\begin{cases}
0, & |g|<2\,\mathrm{se},\\[2pt]
\dfrac{\sigma_0^{2}}{\sigma_0^{2}+\mathrm{se}^{2}}\,g,
& \text{otherwise,}
\end{cases}
\label{eq:live-offset}
\end{equation}
and retain only downward corrections, so a positive live residual never
raises the predicted recovery. The $2\,\mathrm{se}$ condition is an empirical
update gate that leaves small discrepancies uncorrected, and the shrinkage
factor gives greater weight to precise live measurements and less to uncertain
ones. We set $\sigma_0=0.023$. This value is the 90th percentile, over 12
fragment-attack cells (the three fragments under JPEG, Crop 75\%, Rotation,
and Regen), of the between-source standard deviation of the cell's mean bit
accuracy. It was estimated in a pilot measurement with 30 images from each of
the five source groups, after subtracting the within-source sampling variance.

\paragraph{Acceptance-Rate Correction.}
The image-level acceptance prediction is derived from stored per-image scores,
so a discrepancy in acceptance rate cannot be added directly as a scalar
offset. Instead, we shift the stored per-image scores of the cell downward by
a common amount. A correction is made only when the predicted rate at the
candidate's strength exceeds the observed rate $\hat p$, and either $\hat p$
is below the required $\rho$ or the excess is at least two standard errors,
with $\mathrm{se}=\sqrt{\max(\hat p(1-\hat p),0.09)/N_a}$. The shift is found
by 16 bisection steps over $[-0.4,0]$: it is the smallest downward shift, to
this resolution, at which the predicted rate no longer exceeds $\hat p$. The
rate is a step function of the shift, so the corrected prediction may fall
below $\hat p$ rather than match it, and a shift of $-0.4$ is applied when no
smaller one suffices.

\paragraph{Request-Local Model Update.}
A live measurement is taken at one strength, whereas the solver reads a curve.
A pointwise update could allow the next solve to move to a neighboring
strength on the same optimistic curve and reproduce the failed configuration.
We therefore model the request-local discrepancy as a constant offset over the
whole strength range of the observed fragment-attack cell. This is a modelling
approximation, since the discrepancy is observed at one strength only. The
mean correction $\Delta b$ is added to the predicted mean recovery of the
cell, and the rate correction is added to its stored per-image scores. The two
corrections are applied independently, and when several failed candidates
yield a correction for the same cell, the most negative one is kept. If the
failed candidate used a geometric recovery stage, the corrections are applied
to that stage's response, since that is the prediction used by the solver.

All corrections are applied to a request-local copy of the Performance
Database and never modify the shared offline measurements. \sys{} then
re-solves the joint configuration problem once using the corrected model. This
solve is not restricted to new structures, and the resulting candidate must
pass the same live validation criteria before it can be accepted.

\paragraph{Margin Fallback.}
A margin level thus validates at most four candidates, three structures and
one refined candidate. If none passes, or if the solver finds no candidate at
that level, \sys{} moves to the next level of a fixed schedule of rate and
mean margins $(\eta_{\mathrm{rate}},\eta)$: $(0.06,0.02)$, $(0.04,0.02)$,
$(0.02,0.01)$, and $(0,0)$, entered at the level of the initial rate margin.
Each level starts again from the uncorrected database, with a new set of up to
three structures and its own refinement, so corrections are not carried across
levels. A request therefore validates at most 16 candidates from an initial
rate margin of 0.06 and at most 12 from 0.04. The request, the threshold
$\tau(c,\alpha)$, the required rate $\rho$, and the PSNR and latency checks are
identical at every level. If the last level ends without an accepted
candidate, \sys{} reports the pipeline-level \textsc{unsat} of
Sec.~\ref{sec:measurement-backed}. This outcome covers both cases: the solver
finds no candidate at any level, or every validated candidate fails.
\subsection{Watermark Deployment}
\label{sec:method-deployment-details}

The composition library contains VINE~\citep{vine},
TrustMark~\citep{trustmark}, and VideoSeal~\citep{videoseal}. Their native
controls are VINE residual scale, TrustMark embedding strength, and VideoSeal
blend scale \rev{(Appendix~\ref{sec:native-strength})}. Every fragment transmits the same 100-bit word: VINE and
TrustMark carry 100 bits natively, whereas VideoSeal's message vector is
longer, so the deployment writes the word into its first 100 positions and
reads back the corresponding 100 logits, leaving the rest unused. Sequential
embedding in the accepted order follows Eq.~\ref{eq:composition}, with $w$ entering
each fragment through that fragment's own key-derived permutation and mask.
Scale and tile modify the embedding step of VINE and TrustMark, respectively,
resync adds its template once after the last fragment, and angle leaves the
embedding unchanged (Appendix~\ref{sec:geometric-stages}).
Characterization and calibration embed under reserved calibration identities,
one per image, and released images are embedded under deployment identities
outside this set, which the FPR guarantee of Appendix~\ref{sec:fpr-selection}
requires. The keyed permutations and masks of all identities are identically
distributed, so the calibration identities are a sample from the same
distribution as a deployment identity. The last paragraph of this subsection
evaluates the selected configurations under fresh identities.

At verification, the decoder reverses each fragment's key-derived permutation
and mask, evaluates the permitted individual readouts and primary fusion, and
then applies the enabled geometric search when needed. All paths use the same
configuration-dependent threshold in Eq.~\ref{eq:unified-threshold}.

\begin{revblock}
\paragraph{Fresh Deployment Identities.}
\label{sec:fresh-identity}
We evaluate whether configurations selected during calibration, under
calibration identities, retain request satisfaction on held-out images under
previously unseen deployment identities. We take
the 689 distinct configurations that \sysfull{} deploys for its 7,130 satisfied
requests and re-evaluate them on 100 images that were not used to select them,
embedding and verifying each image under an identity that appears nowhere in
characterization or calibration. Attacks, thresholds, acceptance criteria, and
code are unchanged. On these images, all 7,130 requests pass under the
calibration identities and 7,120 pass under fresh identities (99.86\%). The 10
failures come from three S5 configurations and a single attack, Rinse-2x, whose
acceptance rates are 0.91 to 0.94 under the calibration identities and 0.87 to
0.89 under fresh ones, just below the required $\rho=0.90$. No configuration
passes only under fresh identities. Across all 8,101 configuration and attack
cells, the acceptance rate is unchanged in 6,574 cells and changes by $-0.03$
percentage points on average. The selected configurations therefore transfer to
fresh identities, apart from a few cells at the acceptance margin under
Rinse-2x.
\end{revblock}

\begin{revblock}
\subsection{Native Strength Controls}
\label{sec:native-strength}

Each fragment exposes one native strength control $\lambda_f$. In all three
methods it scales the watermark residual that the pretrained encoder adds to
the image, and no network is retrained. For VINE, the encoder runs at
$256\times256$, and we add $\lambda$ times its bicubically upsampled residual
to the original image. For TrustMark, $\lambda$ is the embedding-strength
argument of the released library, which multiplies the upsampled encoder
residual before it is added to the cover. For VideoSeal, $\lambda$ multiplies
the watermark blend scale of the released model. In each case $\lambda=1$ is
the released operating point.

Table~\ref{tab:native-strength} lists where each control is characterized.
Recovery and distortion are measured at the listed knots, on 100 images per
knot. Attacks whose profiling is costly use coarser grids within the same
range. Image editing is profiled at nine evenly spaced strengths per fragment.
UnMarker is profiled on 30 images, at nine knots for VINE (0.2 to 1.0 in steps
of 0.1), all eleven for TrustMark, and twelve for VideoSeal, and CtrlRegen$+$
uses 50 images at some knots. The solver treats $\lambda_f$ as a continuous variable.
Its upper bound is the largest knot. Its lower bound is the larger of the
smallest knot and a clean-image minimum strength, below which the fragment is
not read reliably even without an attack. This minimum depends on the FPR
budget only, through the single-test threshold $\theta_1(\alpha)$, which is
Eq.~\ref{eq:unified-threshold} with $T=1$. It does not depend on $T(c)$ or on
the enabled stage. It equals the lower value given in the table when
$\theta_1\le63$ and the higher value when $\theta_1\ge83$, and it is
interpolated linearly in $\theta_1$ in between. For the six FPR budgets of the
benchmark, $\theta_1$ is 57, 63, 69, 74, 80, and 83, so the minimum lies at
fractions 0, 0, 0.30, 0.55, 0.85, and 1 of the way from the lower value to the
higher one. Both values come from a clean-image sweep on 100
characterization images: the lower value is the smallest strength at
which at least 99\% of the images reach 63 matching bits, and the higher value
is the smallest strength at which the mean bit accuracy minus 2.33 standard
deviations is at least 0.83. Within these bounds every response is
interpolated linearly between neighbouring knots (Eq.~\ref{eq:pwl}). VINE is
thus only attenuated relative to its released strength, whereas TrustMark and
VideoSeal may also be amplified.

The three controls are native to their methods and are not on a common scale,
so equal values of $\lambda$ do not imply equal distortion or robustness
across fragments. Only the measured response curves relate a strength to
recovery and distortion.

The compared methods use these controls differently. The fixed
single-watermark baselines and their Geo. variants embed at the released
setting $\lambda=1$. Optimized Single and Always All Three, the two composition
controls of Table~\ref{tab:main}, run the same solver as \sys{} with the
fragment subset restricted to a single fragment or fixed to all three, and they
optimize strengths continuously over the same ranges. The greedy variant also
optimizes strengths over these ranges.
\end{revblock}

\begin{table}[!htbp]
\centering
\revcolor
\caption{Native strength controls of the three fragments.}
\label{tab:native-strength}
\papertablestyle
\begin{tabularx}{\linewidth}{@{}l X@{}}
\toprule
\multicolumn{2}{@{}l}{\textit{VINE}} \\
Control & scale of the encoder residual added to the image \\
Knots & 0.14, 0.17, 0.2, 0.3, 0.4, 0.5, 0.6, 0.7, 0.8, 0.9, 1.0 \\
Range & 0.14 to 1.0; the lower end rises to 0.30 at the strictest FPR budget \\
\midrule
\multicolumn{2}{@{}l}{\textit{TrustMark}} \\
Control & library embedding strength, a scale of the encoder residual \\
Knots & 0.4, 0.55, 0.7, 0.85, 1.0, 1.15, 1.3, 1.45, 1.6, 1.8, 2.0 \\
Range & 0.55 to 2.0; the lower end rises to 0.70 at the strictest FPR budget \\
\midrule
\multicolumn{2}{@{}l}{\textit{VideoSeal}} \\
Control & multiplier of the released watermark blend scale \\
Knots & 0.5, 0.625, 0.75, 0.875, 1.0, 1.125, 1.25, 1.375, 1.5, 1.75, 2.0, 2.25, 2.5, 2.8, 3.2 \\
Range & 0.5 to 3.2; the lower end rises to 0.625 at the strictest FPR budget \\
\bottomrule
\end{tabularx}
\end{table}

%
\ifdefined\lemma\else\newtheorem{lemma}{Lemma}\fi
\ifdefined\proposition\else\newtheorem{proposition}{Proposition}\fi
\ifdefined\corollary\else\newtheorem{corollary}{Corollary}\fi

\begingroup\revcolor
\section{FPR Control under Request-Local Selection}
\label{sec:fpr-selection}

Live calibration evaluates candidate configurations and may apply
request-local corrections, refinement, or fallback before selecting one final
configuration $c_{\mathrm{out}}$. We prove that the selected verifier preserves the
configuration-level FPR budget in the fresh-null experiment defined below.

\paragraph{Keyed alignment and verification.}
Let $\kappa$ be the master secret and $F_\kappa$ a keyed pseudorandom
function (PRF). Distinct, unambiguous input labels encode the watermark fragment $f$,
identity $v$, output purpose (mask or permutation), and counter. For each
fragment and identity, the derived sign mask and permutation are
\begin{equation}
 M_{f,v}\in\{-1,+1\}^{n},\qquad
 \sigma_{f,v}\in\mathfrak S_n,\qquad n=100.
\end{equation}
The mask is obtained from distinct parity bits of the PRF stream, and the
permutation from a separate stream. The target codeword $w_v\in\{0,1\}^n$
is a public function of $v$ (or, more generally, is fixed independently of
the target mask stream). On a candidate view, the fragment decoder outputs
raw logits $y_f\in\mathbb R^n$, and verification aligns them as
\begin{equation}
 \ell_{f,j}
 =M_{f,v}[\sigma_{f,v}(j)]\,y_f[\sigma_{f,v}(j)],
 \quad j\in\{1,\ldots,n\}.
 \label{eq:selection-aligned-llr}
\end{equation}
For a selected fragment set $S$, the primary fused readout is
$\ell_{\mathrm{fuse},j}=\sum_{f\in S}\ell_{f,j}$.
For any individual or fused test $r$ with aligned readout $\ell_r$,
the match count is
\begin{equation}
 K_r=\sum_{j=1}^{n}
 \mathbf 1\!\left[(2w_{v,j}-1)\ell_{r,j}>0\right].
 \label{eq:selection-match-count}
\end{equation}
In particular, a zero logit is a mismatch.

\paragraph{Selection history and candidate tests.}
Let $u$ denote the deployment request, and let $V_{\mathrm{cal}}$ contain
all identities used for offline characterization and live calibration.
The transcript $\mathcal H$ comprises every observation and random choice
available to the \emph{configuration-selection algorithm}, including database
measurements, live outcomes, refinements, and fallbacks. The algorithm does not
read $\kappa$ directly: its key-dependent operations use only identities in
$V_{\mathrm{cal}}$. The deployed configuration is
\begin{equation}
 c_{\mathrm{out}}=A(u,\mathcal H),
 \label{eq:selection-output}
\end{equation}
where $A$ is an arbitrary selection rule and may examine any number of
candidate configurations. The deployment identity $v^*$ is fixed before
selection or chosen afterwards from public information and $\mathcal H$, and
its choice does not depend on any PRF output indexed by $v^*$. The key
material of $v^*$ is derived only after $c_{\mathrm{out}}$ is frozen, when the
protector embeds the released images, and it never enters $\mathcal H$.

For a configuration $c$ and candidate image $x$, define
$\mathcal R(c,x)$ to be a \emph{mask-independent potential set} containing
all individual, primary fused, and enabled geometric verification tests.
It is determined from $(c,x)$ and randomness independent of the target
key material, before testing whether any path matches. Early stopping and
fallback may choose which tests are actually executed, but those executed
tests form a subset of $\mathcal R(c,x)$. Assume
\begin{equation}
 |\mathcal R(c,x)|\le T(c),
 \label{eq:selection-path-budget}
\end{equation}
where $T(c)$ is the verification-test budget of
Eq.~\ref{eq:unified-budget}, which sets the paper's FPR threshold. The candidate set may include data-dependent geometric views,
provided their generation does not depend on the target key, identity-specific
secret outputs, or match feedback.

\paragraph{Assumptions.}
We use the following conditions.
\begin{itemize}
\item[(P)] \textbf{Ideal PRF.} In the information-theoretic hybrid,
$F_\kappa$ is replaced by a uniformly random function. Outputs at distinct
input labels are independent and uniform. The label design ensures that
mask bits for different fragments/identities/positions, and permutation
streams, use disjoint inputs.

\item[(F)] \textbf{Fresh deployment identity.} The deployed identity
$v^*\notin V_{\mathrm{cal}}$ is not used during characterization or
calibration: the selection transcript $\mathcal H$ is measurable with respect
to PRF outputs on non-$v^*$ labels and target-independent randomness. The
protector may derive the key material of $v^*$ once $c_{\mathrm{out}}$ is
frozen, in order to embed the released images. This use lies outside
$\mathcal H$, and (N) keeps it away from the null query.

\item[(N)] \textbf{Fresh null query.} The negative image $X^-$ and any
information used to generate or select it are independent of the PRF outputs
at $v^*$ labels, conditional on $(u,\mathcal H,v^*)$. The image may be
never-watermarked or watermarked under another independent identity. This
assumption excludes images derived from target-$v^*$ watermarked releases
and samples selected using verification feedback for $v^*$.

\item[(R)] \textbf{Key-independent readouts and view generation.} For
fixed $(c,x)$, the raw decoder outputs for every potential test, the
potential test set $\mathcal R(c,x)$, and their decoder-side randomness do
not depend on the secret PRF outputs associated with the queried target
identity. The target key material is used only through the alignment in
Eq.~\ref{eq:selection-aligned-llr}; the primary fusion is a linear sum of
aligned logits. The threshold depends on $(c,\alpha)$ but not on the
target masks or target verification feedback.
\end{itemize}

\begin{lemma}[Fresh target material]
\label{lem:selection-fresh-material}
Under (P), (F), and (N), conditional on $(u,\mathcal H,v^*,X^-)$, the
family $(M_{f,v^*},\sigma_{f,v^*})_f$ is distributed as freshly derived
ideal-function material. In particular, the target masks have mutually
independent uniform entries and are independent of the target permutations.
\end{lemma}
\begin{proof}
In the ideal hybrid, the outputs of a uniformly random function at the labels
of $v^*$ are uniform and independent of its outputs at all other labels,
however adaptively those other labels are queried. By (F), the transcript
$\mathcal H$ is a function of outputs at non-$v^*$ labels and
target-independent randomness, and by (N), the negative image and its
generation carry no information about the outputs at $v^*$ labels. Embedding
released images under $v^*$ after selection changes neither fact, because
those images enter neither $\mathcal H$ nor $X^-$. Conditioning on the choice
of $v^*$ does not change the distribution either, because that choice does
not depend on the outputs at its labels. Domain-separated
labels then give independent uniform mask words, hence independent fair mask
signs, and an independent permutation stream. Only the bijectivity and
mask-independence of the permutation are used below; exact uniformity of a
practical permutation algorithm is unnecessary.
\end{proof}

\begin{lemma}[Masked-null domination]
\label{lem:selection-masked-null}
Assume (P), (F), (N), and (R). For each potential verification test
$r\in\mathcal R(c_{\mathrm{out}},X^-)$ and every integer $h$, we have
\begin{equation}
\begin{aligned}
 &\Pr\!\left[K_r\ge h\mid u,\mathcal H,v^*,X^-\right]
 \le \Pr[B\ge h],\\
 &B\sim\operatorname{Bin}(n,1/2).
\end{aligned}
 \label{eq:selection-masked-null}
\end{equation}
This guarantee does not require independence among different tests or among
the raw decoder bits.
\end{lemma}
\begin{proof}
Condition on $(u,\mathcal H,v^*,X^-)$, on the decoder-side randomness for
all potential tests, and on all target permutations. The selected
configuration $c_{\mathrm{out}}$, all raw readouts, the candidate test set, and the
codeword $w_{v^*}$ are now fixed by (R). By
Lemma~\ref{lem:selection-fresh-material}, every target mask remains a vector
of independent fair signs, independent across fragments.

\emph{Individual readout.}
For fragment $f$ and bit position $j$, define
$I_{f,j}=\mathbf 1[(2w_{v^*,j}-1)\ell_{f,j}>0]$.
If the relevant raw logit is zero, $I_{f,j}=0$. Otherwise its sign is fixed,
and multiplication by the independent fair mask yields
$\Pr[I_{f,j}=1]=1/2$. Because $\sigma_{f,v^*}$ is a bijection,
distinct positions use distinct independent mask entries. Hence the
indicators are independent across $j$, with success probabilities at most
$1/2$.

\emph{Fused readout.}
At position $j$, write
\begin{equation}
 \varepsilon_j
 =\left(M_{f,v^*}[\sigma_{f,v^*}(j)]\right)_{f\in S},
 \qquad
 \ell_{\mathrm{fuse},j}=g_j(\varepsilon_j).
\end{equation}
The vector $\varepsilon_j$ is uniform on $\{-1,+1\}^{|S|}$, and the
fused-logit rule is odd:
$g_j(-\varepsilon)=-g_j(\varepsilon)$.
Consequently, the events
$(2w_{v^*,j}-1)g_j(\varepsilon_j)>0$ and
$(2w_{v^*,j}-1)g_j(\varepsilon_j)<0$ have equal probability. Since zero
logits do not match, the match probability is at most $1/2$.
Different positions use disjoint mask entries in every fragment, so their
indicators are independent across $j$.

\emph{Stochastic domination.}
For either type of test, $K_r$ is a sum of independent Bernoulli indicators
with success probabilities $p_j\le1/2$. Using independent uniforms
$U_j\sim\operatorname{Uniform}(0,1)$, couple each indicator as
$\mathbf 1[U_j<p_j]$ with $\mathbf 1[U_j<1/2]$. The former is no larger
than the latter, and the sum of the latter has distribution
$\operatorname{Bin}(n,1/2)$. Therefore
$\Pr[K_r\ge h]\le\Pr[B\ge h]$ for fixed decoder randomness and
permutations. Averaging over them preserves the bound.
\end{proof}

\begin{proposition}[FPR control after adaptive selection]
\label{prop:fpr-selection}
Suppose (P), (F), (N), and (R) hold and that deployment uses only the
final frozen configuration $c_{\mathrm{out}}=A(u,\mathcal H)$. Let
$\theta(c,\alpha)\in\{0,1,\ldots,n+1\}$ be a threshold, such as that of
Eq.~\ref{eq:unified-threshold}, satisfying
\begin{equation}
 T(c)\Pr[B\ge\theta(c,\alpha)]\le\alpha,
 \qquad B\sim\operatorname{Bin}(n,1/2),
 \label{eq:selection-threshold}
\end{equation}
where the sentinel threshold $n+1$ means acceptance is impossible.
Then for every selection algorithm $A$ and every
$\alpha\in(0,1)$,
\begin{equation}
 \boxed{
 \Pr_0\!\left[\operatorname{Accept}(X^-;c_{\mathrm{out}},v^*)
       \mid u,\mathcal H,v^*\right]\le\alpha .}
 \label{eq:selection-fpr-guarantee}
\end{equation}
Consequently $\Pr_0[\operatorname{Accept}(X^-;c_{\mathrm{out}},v^*)]\le\alpha$.
The bound is independent of the number of candidate configurations examined
during calibration.
\end{proposition}
\begin{proof}
Condition on $(u,\mathcal H,v^*,X^-)$ and on the target-independent
randomness used to form the candidate view set. This fixes $c_{\mathrm{out}}$,
$\theta=\theta(c_{\mathrm{out}},\alpha)$, and $\mathcal R(c_{\mathrm{out}},X^-)$. The verifier can
accept only if an actually executed test reaches the threshold, and every
executed test belongs to the potential set. Thus, by the union bound and
Lemma~\ref{lem:selection-masked-null},
\begin{align}
 &\Pr_0\!\left[\operatorname{Accept}(X^-;c_{\mathrm{out}},v^*)
       \mid u,\mathcal H,v^*,X^-\right] \nonumber\\
 &\quad\le
 \sum_{r\in\mathcal R(c_{\mathrm{out}},X^-)}
 \Pr_0\!\left[K_r\ge\theta
       \mid u,\mathcal H,v^*,X^-\right] \nonumber\\
 &\quad\le |\mathcal R(c_{\mathrm{out}},X^-)|\Pr[B\ge\theta] \nonumber\\
 &\quad\le T(c_{\mathrm{out}})\Pr[B\ge\theta] \le\alpha.
 \label{eq:selection-union-bound}
\end{align}
For the sentinel $\theta=n+1$, all tests fail deterministically.
Averaging over $X^-$ and the target-independent candidate-generation
randomness yields Eq.~\ref{eq:selection-fpr-guarantee}; averaging over
the selection transcript, request, and permitted identity choice yields the
unconditional guarantee. No independence between geometric views is needed.
The bound depends only on the verification paths of the final deployed
configuration.
\end{proof}

\begin{corollary}[Computational FPR guarantee]
\label{cor:selection-prf}
Suppose HMAC-SHA512 with secret key $\kappa$ replaces the ideal random
function and is secure as a pseudorandom function. For any efficient
implementation of the complete characterization, calibration, null-query
generation, and verification experiment making at most $Q_{\mathrm{PRF}}$
PRF queries and running in time at most $t_{\mathrm{adv}}$ (including
simulation overhead),
\begin{equation}
\begin{aligned}
 &\Pr_{0,\mathrm{real}}[\operatorname{Accept}(X^-;c_{\mathrm{out}},v^*)]\\
 &\quad\le\alpha+\operatorname{Adv}^{\mathrm{PRF}}_{\mathrm{HMAC\text{-}SHA512}}(t_{\mathrm{adv}},Q_{\mathrm{PRF}}).
\end{aligned}
 \label{eq:selection-prf-bound}
\end{equation}
\end{corollary}
\begin{proof}
Construct a PRF distinguisher that runs the full experiment with oracle
access to a function $O$, uses $O$ for every key-derivation query, and
outputs one exactly when the final verifier accepts the fresh null image.
If $O$ is an ideal random function, the output probability is at most
$\alpha$ by Proposition~\ref{prop:fpr-selection}. If $O$ is HMAC-SHA512
under a uniformly sampled secret key, the output probability is the real
false-acceptance probability. Their difference is bounded by the
distinguisher's PRF advantage, proving the claim.
\end{proof}

\paragraph{Scope.}
Under assumptions (P), (F), (N), and (R), the guarantee applies to fresh-null
verification with fresh deployment identities; its probability is taken over
random key generation (or random-function outputs in the ideal hybrid).
\endgroup

\section{Geometric Recovery Stages}
\label{sec:geometric-stages}

\begin{figure*}[!htbp]
\centering
\includegraphics[width=0.8\linewidth]{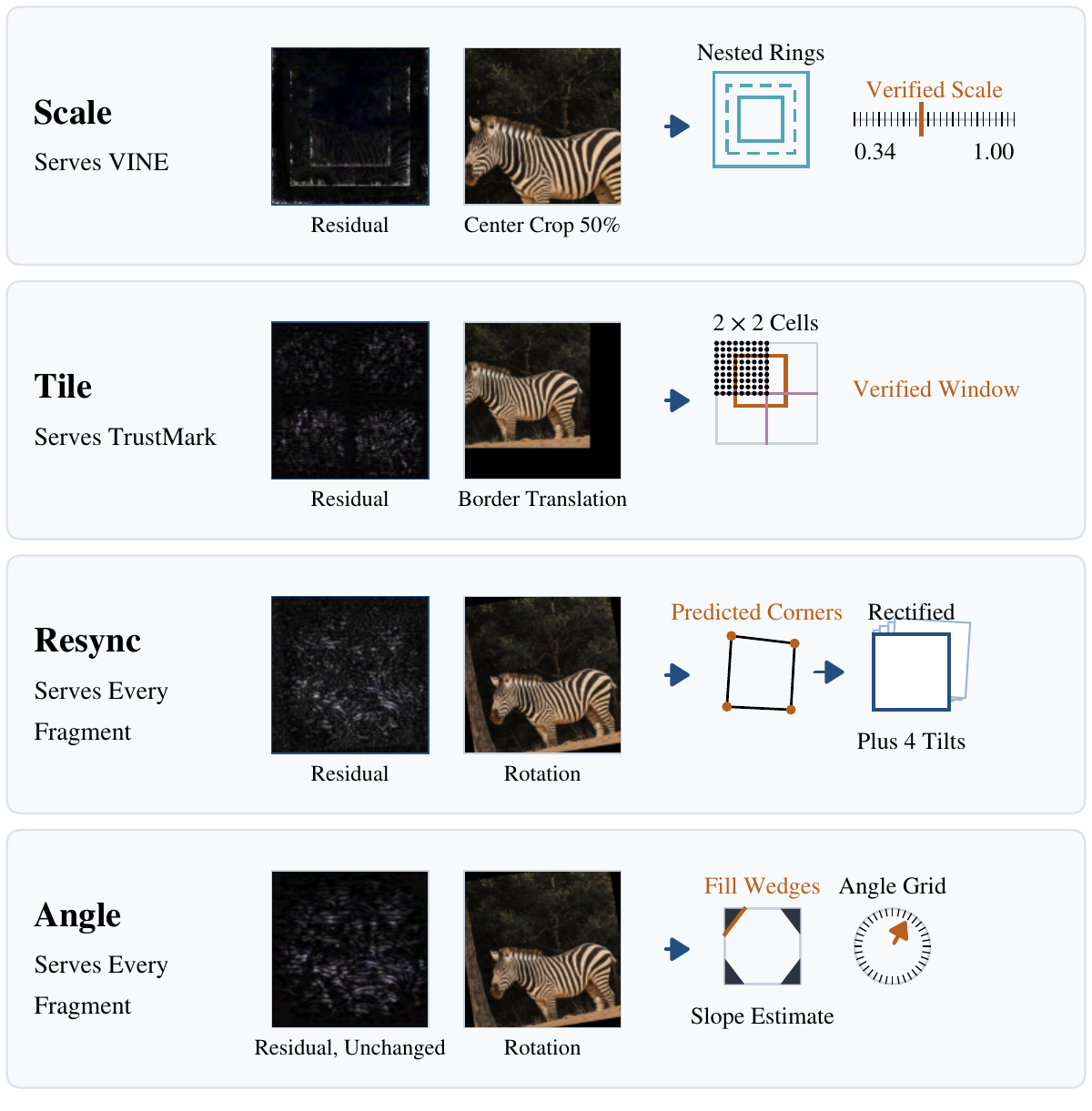}
\caption{The four optional geometric stages, with the embedded residual each one produces and the views it searches at decoding.}
\label{fig:geometric-stages}
\end{figure*}

Geometric transformations affect the three fragments differently, so we first
probe where the recoverable signal of each fragment lies and then describe the
four optional recovery stages.
Figure~\ref{fig:geometric-stages} shows the stages: the residual each
one writes into the image, the attack it addresses, and the views it enumerates
at decoding; its residuals use a $12\times$ display gain. The paragraphs below
give the mechanism and the parameters. At most one stage is enabled in a
configuration.

\subsection{Where Each Fragment Writes Its Payload}
\label{sec:payload-locality}

We revert part of a watermarked image to the cover and read the fragment from
the result. Reverting the interior leaves only a border ring, the outer tenth of
the half-width; reverting that ring leaves everything else. The probe uses 24
MS-COCO images of the live slice, each watermarked by one fragment at unit
strength.

\begin{table}[!htbp]
\centering
\caption{Where each fragment writes its payload.}
\label{tab:payload-locality}
\papertablestyle
\begin{adjustbox}{max width=\linewidth}
\begin{tabular}{@{}l ccc c@{}}
\toprule
\toprule
& \multicolumn{3}{c}{\textit{Bit accuracy after reverting}} & \textit{Absolute residual} \\
\cmidrule(lr){2-4} \cmidrule(lr){5-5}
Fragment & Nothing & Interior & Border ring & in the border ring (\%) \\
\midrule
VINE & 1.000 & \textbf{1.000} & 0.482 & \textbf{50.7} \\
TrustMark & 1.000 & 0.500 & \textbf{1.000} & 7.1 \\
VideoSeal & 0.998 & 0.652 & 0.985 & 15.7 \\
\bottomrule
\bottomrule
\end{tabular}
\end{adjustbox}
\end{table}

Table~\ref{tab:payload-locality} reports the result, and VINE and TrustMark come
out opposite.
VINE reads perfectly from the border ring alone and at chance from the interior
alone, and half of its absolute residual lies in that ring, which is a fifth of
the area. These measurements indicate that VINE's recoverable signal is
concentrated near the image border.
TrustMark is the reverse: it reads perfectly from the interior and at chance
from the ring, so it relies on the interior.
VideoSeal lies between the two in this probe. It reads almost perfectly from
the interior and above chance from the ring.

The behaviour of the standalone fragments under geometric attacks is
consistent with this probe. On the frozen measurements at
$\alpha=10^{-4}$, VINE alone accepts no centred crop, no rotation, and no flip,
while still accepting every image under JPEG, as expected for a signal tied to
fixed image coordinates near the border.
TrustMark alone accepts a crop to 75\% of the area, but its acceptance drops to
1\% under the Border-20 translation and to 4\% under a 9 degree rotation.

\subsection{Failure-Driven Geometric Recovery}

\paragraph{Scale, for centred crops of a border signal.}
The scale stage embeds the VINE codeword three times within the VINE step of
the configured order: on the full frame, on its central 384-pixel crop, and on
its central 256-pixel crop, each at strength $\lambda_{\mathrm{VINE}}$ and
pasted back in place. We index a copy by its side factor
$K\in\{1.00,0.75,0.50\}$. A centred crop of side factor $s$ that is resized to
the original size moves the border of copy $K$ to $K/s$ and removes every copy
with $K>s$. A copy with $K\le s$ thus stays inside the frame and offers a
further opportunity for recovery under centred crops. Under a crop to half the
area, only the innermost copy remains, at scale
$0.50/\sqrt{0.5}\approx0.71$. The smallest side factor is 0.50 because VINE
encodes at $256\times256$, so the 256-pixel crop of a $512\times512$ image is
embedded without resampling.
The decoder reads VINE on 133 centred crops whose side factors run from 0.34
to 1.00 in steps of 0.005, and it keeps the first view that verifies. The grid
is dense because a copy is read only in a narrow range of scales. In a scan of
12 images over this grid, mean bit accuracy is at least 0.999 at a copy's own
scale, between 0.36 and 0.81 one grid step away, and at most 0.51 two steps
away. The same scan finds the copies near the predicted scales under the two
benchmark crops, at 0.58 and 0.87 for Crop 75\% and at 0.71 for Crop 50\%,
and no view below 0.50 reaches a mean bit accuracy above 0.54. The lower end
of the range, 0.34, is an implementation setting.

\paragraph{Tile, for translations of an interior signal.}
The tile stage replaces the plain TrustMark embedding within the TrustMark
step: the frame is divided into a $2\times2$ grid of 256-pixel cells, and each
cell is encoded separately with the same codeword at strength
$\lambda_{\mathrm{TrustMark}}$. Repeating the codeword provides several
spatial opportunities for recovery, since a window that is aligned with a
surviving cell carries a complete codeword. The decoder evaluates 81 windows
of $256\times256$ pixels at a stride of 32 pixels and verifies each one, a
finite window search. The translation in our benchmark is Border-20, a
shift of 102 pixels.

\paragraph{Resync, for a warped frame.}
Rotations and projective transformations can disrupt the alignment on which
the decoders rely. The resync stage embeds a SyncSeal
template~\citep{syncseal} once, after all selected fragments, at the fixed
strength of the released model. At decoding, the SyncSeal detector predicts
the four corners of the original frame, and the decoder applies the
corresponding perspective transformation and reads every selected fragment
from the rectified frame. It repeats the reading at tilts of 3 and 6
degrees either way to absorb the residual error of the corner prediction, five
views per fragment in all. The shared rectification serves all selected
fragments.

\paragraph{Angle, for rotation without a template.}
An in-place rotation pads the corners with fill, and the boundary between fill
and content is a straight line whose slope gives the rotation angle. The angle
stage reads this evidence from the image itself. It leaves the embedding
unchanged and does not depend on which fragments a request selected. When the
corners contain fill, the decoder fits this boundary in the two upper corners,
which yields at most two angle magnitudes, and it proposes each magnitude with
both signs and with offsets of 0, $\pm3$, $\pm6$, and $\pm9$ degrees, at most
28 candidates. It then adds a fixed grid from $-180$ to $180$ degrees in steps
of 10 degrees, 37 candidates. Duplicates are removed, and every selected
fragment is read on each candidate view in this order, at most 65 views per
fragment. The 10 degree grid is a fallback for images without usable corners.

\paragraph{Scope.}
Scale and tile change how one fragment is written, so each applies only to that
fragment and only that fragment is read from their views. Resync and angle
restore the frame, so every selected fragment is read from each of their views.
This is the origin of the per-stage test counts in
Eq.~\ref{eq:unified-budget}: the two frame-restoring stages scale with the
number of selected fragments, the two fragment-specific stages do not. The
embedding cost of the enabled stage enters the distortion model once, through
$D_\varphi(c)$, and its decoding overhead enters the latency model once
(Appendix~\ref{sec:method-table-details}). These recovery stages use finite
searches tailored to the specified geometric transformations.

\section{Evaluation Details}
\label{sec:eval-details}

\subsection{Image Pool and Measurement Roles}
\label{sec:eval-datasets-details}

Images are drawn from five source groups:
\begin{itemize}
    \item \textbf{MS-COCO 2017~\citep{coco}:} real photographs with diverse
    native resolutions.

    \item \textbf{DiffusionDB~\citep{diffusiondb}:} SD-1.5 generated images
    with native sides between 512 and 1024 pixels.

    \item \textbf{UltraEdit~\citep{ultraedit}:} real images and image-editing
    examples at $512\times512$.

    \item \textbf{DALL$\cdot$E~3 / SDXL~\citep{sdxl}:} generated images at
    $1024\times1024$.

    \item \textbf{DIV2K~\citep{div2k} and LSDIR~\citep{lsdir}:}
    high-resolution real images ranging from $1024$ pixels to approximately
    2K resolution.
\end{itemize}

The main experiments use three fixed, mutually disjoint slices of 100 images,
each containing 25, 25, 25, 15, and 10 images from the five source groups
above. The slices are not drawn at random: the pool lists the images of each
source in a fixed order, and a slice takes consecutive images of every source
from a common offset, which is 0, 100, and 300 for the three slices. In all
three slices, the ten high-resolution images are consequently DIV2K images,
and the fifteen images of the fourth group come from one public collection of
generated images, which consists primarily of DALL$\cdot$E~3 outputs and does
not label the generator of each image. Every image is converted to RGB and
resized to $512\times512$ with bicubic interpolation and without cropping, so
the aspect ratio of non-square images is not preserved.

Two slices are used exclusively for offline characterization: the first for
fragment strength sweeps, signal-processing and geometric attacks, image
editing, pairwise interaction measurements, distortion modeling, and geometric
recovery stages, and the second for the standalone responses under
regeneration, rinse, and CtrlRegen$+$.
UnMarker is characterized on 30 images of the first slice. The third slice
serves as the live set. It plays the role of the user's images: candidates are
calibrated on it, and the outcome of every request is reported on it, for all
requests and methods. Reported satisfaction is therefore request satisfaction
on this live batch. The separation enforced here is between offline
characterization and the live set, not between calibration and reporting.
Appendix~\ref{sec:method-deployment-details} re-evaluates the deployed
configurations on 100 further images. File-, byte-, and pixel-level digest
checks confirm that no image is shared between slices. Within the live slice,
two DiffusionDB images are identical files, so it contains 99 distinct images.

\subsection{Request Distribution and Attack Settings}
\label{sec:solver-eval-details}
\label{sec:deployment-eval-details}

We construct deployment requests over attacks with different adversarial
capabilities. In all settings, the adversary has access to the released
watermarked image $x_w$, but not to the protector's secret key or the original
image $x_o$. We group additional capabilities into three levels:
\emph{Image Manipulation}, covering standard image transformations;
\emph{Public-Model Access}, which additionally permits the use of publicly
available pretrained models; and \emph{Optimization-Based Removal}, which
allows iterative optimization directly on the released image.

The watermark-removal scenario (S4) includes CtrlRegen+ at strength $0.3$ or
$0.5$, optionally combined with UnMarker. Both target watermark removal but use
different mechanisms: CtrlRegen+ performs controlled regeneration, whereas
UnMarker applies iterative optimization.

\paragraph{Request Composition.}
The main benchmark contains 7,321 distinct deployment requests spanning 20
attack settings and 240 unique attack sets. S4 contributes 822 requests,
including 548 using CtrlRegen+ at strength $0.3$ and 274 at strength $0.5$;
262 of these requests additionally include UnMarker. All compared methods are
evaluated on the same request IDs. Table~\ref{tab:request-set} summarizes the
request distribution across scenarios and FPR budgets, with the corresponding
quality and latency budgets specified in the table note.

\begin{revblock}
\paragraph{Scenario Construction.}
A request is a tuple of an attack set, an FPR budget, a PSNR floor, and a
latency budget. Each scenario fixes the attacks that all of its requests
contain and the attacks that a request may add. All five scenarios require the
five signal-processing attacks: Brightness, Contrast, JPEG Q25, Gaussian blur,
and Gaussian noise. Table~\ref{tab:scenario-rules} gives the remaining rules.
A request contains every required attack of its scenario, exactly one attack
from each group of alternatives, and any subset of the optional attacks. S3 and
S5 thus contain either Regen or Image editing, and S4 contains one CtrlRegen+
strength. Each optional attack appears in 46\% to 52\% of its scenario's
requests, except UnMarker, which appears in 32\% of S4.

The number of attack sets of a scenario follows from these rules: two to the
power of its number of optional attacks, doubled for each group of
alternatives. This gives $2^5=32$ sets in S1 and in S2, $2\cdot2^3=16$ in S3,
$2\cdot2^4=32$ in S4, and $2\cdot2^6=128$ in S5. No set belongs to two
scenarios, and every one of them occurs in the benchmark, which yields the 240
distinct attack sets. The 20 attack settings of
Table~\ref{tab:attack-baselines} are exactly the attacks named by these rules.

The requests are sampled rather than enumerated: the full product of the 240
attack sets with the budget values of Table~\ref{tab:request-set} would
contain 28,800 tuples, of which the benchmark retains 7,321 distinct ones:
1,551 in S1, 1,532 in S2, 1,373 in S3, 822 in S4, and 2,043 in S5. One attack
set therefore serves about 30 requests on average, which differ in their FPR
budget, PSNR floor, or latency budget. The request list is fixed and is
identical for all methods.
\end{revblock}

\begin{table*}[!htbp]
\centering
\revcolor
\caption{Construction rules of the five scenarios.}
\label{tab:scenario-rules}
\papertablestyle
\begin{tabularx}{\linewidth}{@{}l >{\raggedright\arraybackslash}p{3.3cm} >{\raggedright\arraybackslash}p{3.0cm} >{\raggedright\arraybackslash}X r r@{}}
\toprule
ID & Required & Exactly one of & Optional (any subset) & Sets & Attacks \\
\midrule
S1 & none & none & Resize 256, Horizontal flip, VAE-B, VAE-C, Crop $+$ JPEG & 32 & 5 to 10 \\
S2 & Crop 75\%, Rotation 9\textdegree & none & Resize 256, Horizontal flip, Crop $+$ JPEG, Crop 50\%, Border-20 & 32 & 7 to 12 \\
S3 & VAE-B, VAE-C & Regen or Image editing & Rinse-2x, Resize 256, Horizontal flip & 16 & 8 to 11 \\
S4 & Regen & CtrlRegen$+$ 0.3 or 0.5 & VAE-B, VAE-C, Rinse-2x, UnMarker & 32 & 7 to 11 \\
S5 & Crop 75\%, Rotation 9\textdegree, VAE-B, VAE-C & Regen or Image editing & Rinse-2x, Crop 50\%, Horizontal flip, Resize 256, Crop $+$ JPEG, Border-20 & 128 & 10 to 16 \\
\bottomrule
\end{tabularx}
\papertablenote{Attacks: number of attacks in one request, including the five signal-processing attacks. Regen appears in 773 requests of S3 and 1,543 of S5, and Image editing in 600 and 500.}
\end{table*}

\begin{table*}[!htbp]
\centering
\caption{Request distribution across scenarios and FPR budgets.}
\label{tab:request-set}
\papertablestyle
\begin{adjustbox}{max width=\linewidth}
\begin{tabular}{@{}ll ccc cccccc@{}}
\toprule
\toprule
& & \multicolumn{3}{c}{\textit{Distinct specifications}} & \multicolumn{6}{c}{\textit{Specifications per FPR budget}} \\
\cmidrule(lr){3-5} \cmidrule(lr){6-11}
ID & Scenario & Count & Share (\%) & Attack sets & $10^{-1}$ & $10^{-2}$ & $10^{-4}$ & $10^{-6}$ & $10^{-9}$ & $2^{-37}$ \\
\midrule
S1 & Signal / re-encoding & 1,551 & 21.19 & 32 & 312 & 441 & 345 & 221 & 120 & 112 \\
S2 & Geometry & 1,532 & 20.93 & 32 & 309 & 434 & 355 & 215 & 114 & 105 \\
S3 & Regeneration / editing & 1,373 & 18.75 & 16 & 277 & 328 & 313 & 223 & 125 & 107 \\
S4 & Watermark removal & 822 & 11.23 & 32 & 362 & 188 & 155 & 91 & 9 & 17 \\
S5 & Broad attack coverage & 2,043 & 27.91 & 128 & 438 & 628 & 489 & 274 & 111 & 103 \\
\midrule
Total & & 7,321 & 100.00 & 240 & 1,698 & 2,019 & 1,657 & 1,024 & 479 & 444 \\
\bottomrule
\bottomrule
\end{tabular}
\end{adjustbox}
\papertablenote{%
PSNR floors in dB are drawn from $\{34,35,36,37,38,40\}$ in S1, $\{32,33,34,35,36,38\}$ in S2 and S5, and $\{30,31,32,33,34,36\}$ in S3 and S4; latency budgets in ms from $\{500,1000,2000,4000\}$ in S1, $\{1000,2000,4000,8000\}$ in S2 and S3, $\{2000,4000,8000\}$ in S4, and $\{4000,8000,16000\}$ in S5.
}
\end{table*}

\begin{table*}[!htbp]
\centering
\caption{The 20 attack settings used in evaluation.}
\label{tab:attack-baselines}
\papertablestyle
\begin{tabular}{@{}
>{\raggedright\arraybackslash}p{3.1cm}
>{\raggedright\arraybackslash}p{2.9cm}
>{\raggedright\arraybackslash}p{\dimexpr\linewidth-6.0cm-4\tabcolsep\relax}
@{}}
\toprule
\toprule
Attack & Capability & Configuration \\
\midrule

\multicolumn{3}{l}{\emph{Photometric and signal processing}} \\
\midrule
Brightness & Image Manipulation & pixel values scaled by $0.2$ \\
Contrast & Image Manipulation & contrast factor $0.2$, a blend toward the mean grey level \\
JPEG Q25 & Image Manipulation & quality $25$ \\
Gaussian blur & Image Manipulation & kernel $5\times5$, $\sigma=1$ \\
Gaussian noise & Image Manipulation & additive, $\sigma=0.05$ on the $[0,1]$ scale, clipped \\

\midrule
\multicolumn{3}{l}{\emph{Geometric transformations}} \\
\midrule
Crop 75\% & Image Manipulation & centre crop keeping $75\%$ of the area, side factor $0.866$; resized back to $512$ with bicubic interpolation \\
Crop 50\% & Image Manipulation & centre crop keeping $50\%$ of the area, side factor $0.707$; resized back to $512$ with bicubic interpolation \\
Rotation 9\textdegree & Image Manipulation & $9^\circ$ in place with bicubic interpolation, corners left as black fill \\
Crop $+$ JPEG & Image Manipulation & centre crop to $75\%$ area as above, then JPEG quality $25$ \\
Resize 256 & Image Manipulation & bicubic downscale to $256$, bicubic resize back to $512$ \\
Horizontal flip & Image Manipulation & horizontal mirroring \\
Border-20 & Image Manipulation & translation crop: drop $20\%$ of the side ($102$ pixels) from the top and left, push the remainder to the origin, leave the vacated strips black; frame size preserved \\

\midrule
\multicolumn{3}{l}{\emph{Neural compression}} \\
\midrule
VAE-B~\citep{bmshj2018} & Public-Model Access & \texttt{bmshj2018-hyperprior}, quality level $3$, one encode-decode pass \\
VAE-C~\citep{cheng2020} & Public-Model Access & \texttt{cheng2020-anchor}, quality level $3$, one encode-decode pass \\

\midrule
\multicolumn{3}{l}{\emph{Regeneration and editing}} \\
\midrule
Regen~\citep{zhao2024regen} & Public-Model Access & SD~2-1, DPMSolver; the latent is noised to timestep $60$ of $1{,}000$ (\texttt{noise\_step}$=60$) and denoised over the last three steps of a 50-step schedule with an empty prompt; one pass \\
Rinse-2x & Public-Model Access & two successive Regen passes \\
CtrlRegen$+$ 0.3~\citep{ctrlregen} & Public-Model Access & released pipeline, 50-step schedule, image-to-image strength $0.3$ ($15$ denoising steps) \\
CtrlRegen$+$ 0.5~\citep{ctrlregen} & Public-Model Access & released pipeline, 50-step schedule, image-to-image strength $0.5$ ($25$ denoising steps) \\
Image editing~\citep{instructpix2pix} & Public-Model Access & InstructPix2Pix, $20$ steps, guidance $7.0$, image guidance $1.5$, five global prompts assigned to the images in rotation \\

\midrule
\multicolumn{3}{l}{\emph{Watermark removal}} \\
\midrule
UnMarker~\citep{unmarker} & Optimization-Based Removal & released two-stage attack: centre crop $0.9$ and resize back, then spectral optimization ($2{,}000$ iterations) and filter-based optimization ($500$ iterations) on the released image \\

\bottomrule
\bottomrule
\end{tabular}
\papertablenote{Each attack is applied once to each watermarked image of a configuration, and the result is shared by all requests that contain the attack. Image editing and CtrlRegen$+$ use a fixed seed, whereas Gaussian noise, Regen, and UnMarker draw fresh noise. The editing prompts are ``make it look like a painting'', ``turn it into winter with snow'', ``make it a sunset scene'', ``give it autumn colors'', and ``make it look vintage''.}
\end{table*}

\subsection{Baseline}
\label{sec:eval-baseline-details}
\label{sec:wm-baselines}

The main comparison of Table~\ref{tab:main} includes the full method, the
greedy variant, the two composition controls Optimized Single and Always All
Three, fixed single-watermark controls, optional-geometry single-watermark
controls, MaskWM, and StegaStamp. Enumeration is a further fixed-strength
control that appears only in this supplement.

\paragraph{Optimized Single and Always All Three.}
Both controls run the complete pipeline of \sysfull{} with one added
constraint on the fragment subset. Optimized Single allows at most one
selected fragment. Always All Three selects all three fragments and therefore
drops the requirement that every selected fragment covers at least one
requested attack. Everything else is shared with \sysfull{}: the Performance
Database, continuous strength optimization, the choice of order and geometric
stage, up to three candidate structures per margin level, request-local
refinement, the margin schedule from an initial rate margin of 0.06, the live
images, and the acceptance criteria.

\paragraph{Greedy.}
The greedy variant \sys-G starts  from the empty fragment subset and irrevocably adds
the fragment with the largest predicted coverage gain. For each trial subset,
the full method's SMT model reoptimizes continuous strengths, embedding order, and
compatible geometry. The first complete candidate satisfying the request is
passed to the common live-validation procedure.

\paragraph{Fixed Single-Watermark Controls.}
VINE, TrustMark, and VideoSeal are evaluated at unit strength without geometry.
Their optional-geometry variants additionally try compatible geometric
stages using the same implementations and FPR accounting as \sys{}.

\paragraph{MaskWM and StegaStamp.}
MaskWM~\citep{maskwm} and StegaStamp~\citep{stegastamp} are evaluated as
fixed external baselines. Both embed the same 100-bit transmitted word and use
the common strict acceptance rule.
StegaStamp's evaluated resize pipeline has mean embedding PSNR of
28.55~dB, below the lowest requested floor of 30~dB, so no request is accepted
and its accepted-request fidelity metrics are undefined.

\paragraph{Enumeration.}
Enumeration is a supplementary fixed-strength control. It is not part of
Table~\ref{tab:main} and appears only in Table~\ref{tab:confidence-intervals}
and in Figures~\ref{fig:qualitative-s2} to \ref{fig:qualitative-s4}.
Enumeration evaluates all seven nonempty subsets of the three-fragment library
at unit strength and fixed embedding order. Enumeration + Geo. extends the same
subset sequence with compatible geometric stages. Each candidate is tested
using the common recovery, image-quality, latency, and FPR criteria.

\subsection{Evaluation Metrics}
\label{sec:eval-metric-details}

We evaluate each method along three dimensions: whether it satisfies the complete
deployment request, the fidelity of the resulting watermarked images, and the
deployment latency required by the accepted configuration.

\paragraph{Request Satisfaction.}
A deployment request is satisfied only if the selected configuration passes all
requested attacks during live evaluation while meeting the specified
image-quality and latency requirements. Each attack is evaluated on the 100
live images, except UnMarker, which is evaluated on the first 30 because of
its cost. We report satisfaction separately for
each scenario and use the unweighted mean of the five scenario rates as the
\textbf{Avg.} metric in Table~\ref{tab:main}.

Table~\ref{tab:confidence-intervals} reports nominal 95\% Wilson confidence
intervals computed over requests, conditional on the fixed image pool and
experimental runs. These intervals use the standard independent-request
approximation. For \sysfull{}, the widest interval is that of S4,
$85.40^{+2.25}_{-2.58}\%$, and the scenario-averaged satisfaction is
$96.21^{+0.48}_{-0.57}\%$.

\begin{table*}[!t]
\centering
\caption{Request satisfaction with 95\% confidence intervals.}
\label{tab:confidence-intervals}
\papertablestyle
\begin{adjustbox}{max width=\linewidth}
\begin{tabular}{@{}l cccccc@{}}
\toprule
\toprule
& \multicolumn{6}{c}{\textit{Request satisfaction (\%) $\uparrow$ with 95\% CI}} \\
\cmidrule(lr){2-7}
Method & S1 & S2 & S3 & S4 & S5 & \textbf{Avg.} \\
\midrule

StegaStamp
& $0.00^{+0.25}_{-0.00}$
& $0.00^{+0.25}_{-0.00}$
& $0.00^{+0.28}_{-0.00}$
& $0.00^{+0.47}_{-0.00}$
& $0.00^{+0.19}_{-0.00}$
& \cellcolor{orange!8}$0.00^{+0.13}_{-0.00}$ \\

MaskWM
& $10.32^{+1.61}_{-1.42}$
& $0.00^{+0.25}_{-0.00}$
& $0.00^{+0.28}_{-0.00}$
& $0.00^{+0.47}_{-0.00}$
& $0.00^{+0.19}_{-0.00}$
& \cellcolor{orange!8}$2.06^{+0.35}_{-0.28}$ \\

VINE
& $8.83^{+1.52}_{-1.31}$
& $0.00^{+0.25}_{-0.00}$
& $42.68^{+2.63}_{-2.59}$
& $54.26^{+3.38}_{-3.42}$
& $0.00^{+0.19}_{-0.00}$
& \cellcolor{orange!8}$21.15^{+0.91}_{-0.90}$ \\

TrustMark
& $100.00^{+0.00}_{-0.25}$
& $0.00^{+0.25}_{-0.00}$
& $23.38^{+2.31}_{-2.16}$
& $0.00^{+0.47}_{-0.00}$
& $0.00^{+0.19}_{-0.00}$
& \cellcolor{orange!8}$24.68^{+0.48}_{-0.44}$ \\

VideoSeal
& $27.98^{+2.29}_{-2.18}$
& $49.74^{+2.50}_{-2.50}$
& $4.01^{+1.17}_{-0.92}$
& $0.00^{+0.47}_{-0.00}$
& $2.74^{+0.80}_{-0.62}$
& \cellcolor{orange!8}$16.89^{+0.74}_{-0.70}$ \\

\midrule
VINE + Geo.
& $8.83^{+1.52}_{-1.31}$
& $0.00^{+0.25}_{-0.00}$
& $42.68^{+2.63}_{-2.59}$
& $62.65^{+3.24}_{-3.36}$
& $0.00^{+0.19}_{-0.00}$
& \cellcolor{orange!8}$22.83^{+0.89}_{-0.89}$ \\

TrustMark + Geo.
& $100.00^{+0.00}_{-0.25}$
& $94.71^{+1.01}_{-1.24}$
& $23.38^{+2.31}_{-2.16}$
& $0.00^{+0.47}_{-0.00}$
& $11.94^{+1.48}_{-1.34}$
& \cellcolor{orange!8}$46.01^{+0.59}_{-0.57}$ \\

\midrule
Optimized Single
& $100.00^{+0.00}_{-0.25}$
& $100.00^{+0.00}_{-0.25}$
& $69.70^{+2.37}_{-2.48}$
& $73.72^{+2.89}_{-3.11}$
& $23.45^{+1.89}_{-1.79}$
& \cellcolor{orange!8}$73.37^{+0.84}_{-0.88}$ \\

Always All Three
& $99.61^{+0.21}_{-0.45}$
& $100.00^{+0.00}_{-0.25}$
& $97.31^{+0.73}_{-1.00}$
& $77.74^{+2.71}_{-2.97}$
& $97.70^{+0.57}_{-0.75}$
& \cellcolor{orange!8}$94.47^{+0.57}_{-0.65}$ \\

\sysgreedy{}
& $100.00^{+0.00}_{-0.25}$
& $97.39^{+0.69}_{-0.92}$
& $87.55^{+1.64}_{-1.85}$
& $76.03^{+2.79}_{-3.04}$
& $94.27^{+0.93}_{-1.09}$
& \cellcolor{orange!8}$91.05^{+0.69}_{-0.77}$ \\

\sysfull{}
& $100.00^{+0.00}_{-0.25}$
& $100.00^{+0.00}_{-0.25}$
& $97.31^{+0.73}_{-1.00}$
& $85.40^{+2.25}_{-2.58}$
& $98.34^{+0.47}_{-0.65}$
& \cellcolor{orange!8}$96.21^{+0.48}_{-0.57}$ \\

\midrule
\multicolumn{7}{@{}l}{\textit{Supplementary fixed-strength controls, not in Table~\ref{tab:main}}} \\
Enumeration
& $100.00^{+0.00}_{-0.25}$
& $85.38^{+1.68}_{-1.86}$
& $88.78^{+1.56}_{-1.78}$
& $55.60^{+3.36}_{-3.41}$
& $51.59^{+2.16}_{-2.17}$
& \cellcolor{orange!8}$76.27^{+0.92}_{-0.96}$ \\

Enumeration + Geo.
& $100.00^{+0.00}_{-0.25}$
& $100.00^{+0.00}_{-0.25}$
& $88.78^{+1.56}_{-1.78}$
& $63.99^{+3.21}_{-3.34}$
& $63.29^{+2.06}_{-2.11}$
& \cellcolor{orange!8}$83.21^{+0.82}_{-0.87}$ \\

\bottomrule
\bottomrule
\end{tabular}
\end{adjustbox}

\papertablenote{Entries are $p^{+\Delta_{\mathrm{upper}}}_{-\Delta_{\mathrm{lower}}}$ in percentage points. For Avg., the lower and upper deviations of the five scenarios are combined as the root sum of squares divided by five.}
\end{table*}

\begin{figure*}[tp]
\centering
\includegraphics[width=0.8\linewidth]{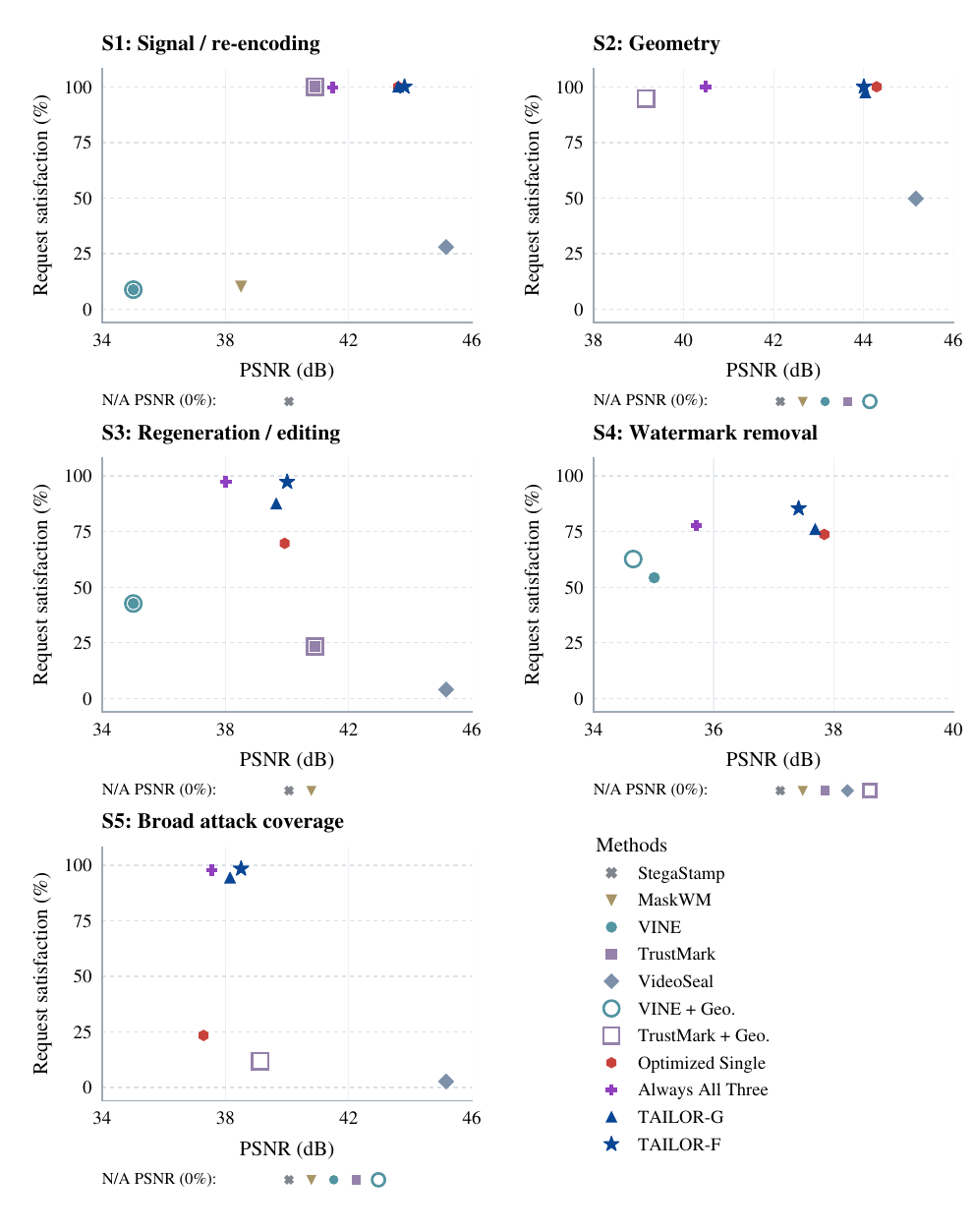}
\caption{Request satisfaction and accepted-request PSNR across five deployment scenarios.}
\label{fig:quality-scenarios}
\end{figure*}

\paragraph{Paired Quality Comparison.}
Because different methods may satisfy different subsets of requests, raw
fidelity averages are not directly comparable. For each baseline $b$, we
therefore compare image fidelity on the requests jointly accepted by $b$ and
\sys{}. The PSNR-P column in Table~\ref{tab:main} reports the mean PSNR of the
baseline and \sys{} on this shared request set, using the same images for both
methods. Each baseline induces a different intersection, so paired comparisons
should be interpreted within each row rather than across rows. Figure~\ref{fig:quality-scenarios} further shows request satisfaction and
accepted-request PSNR across the five deployment scenarios, providing a
scenario-level view of the coverage-fidelity trade-off.

\paragraph{Latency Accounting.}
We distinguish the predicted latency from the measured deployment latency.
The predicted latency $L(c)$ of Appendix~\ref{sec:method-table-details} is
built from offline measurements and is used by the solver during selection.
The measured deployment latency of a configuration is its embedding and
verification time under the requested attacks. Timing measurements use warm,
resident models at batch size one on NVIDIA A100-SXM4-80GB GPUs with CUDA
synchronization, with three repetitions per image after three warm-up calls.
For configuration $c$ and deployment request
$u=(\mathcal A_u,\alpha,q,t)$, the measured latency is
\begin{equation}
L_{\mathrm{meas}}(c,u)
=
\max_{a\in\mathcal A_u}
\frac{1}{N_a}
\sum_{i=1}^{N_a}
\left(
t_i^{\mathrm{embed}}(c)
+
t_{i,a,\alpha}^{\mathrm{decode}}(c)
\right),
\label{eq:deployment-latency}
\end{equation}
where $\mathcal A_u$ is the set of attacks requested by $u$, $N_a$ is the
number of evaluated images under attack $a$, $t_i^{\mathrm{embed}}(c)$ is the
time required to embed configuration $c$ into image $i$, and
$t_{i,a,\alpha}^{\mathrm{decode}}(c)$ is the verification time for the resulting
image under attack $a$ and FPR budget $\alpha$. The measured latency is thus
the maximum, over the requested attacks, of the mean embedding and
verification time per image. It is a mean over images and not a per-image
worst case.
Model loading, image I/O, and attack execution are excluded from the latency
budget.

\section{Additional Ablation Studies}
\label{sec:eval-more-results}
\label{sec:eval-ablation-details}

We examine the screening margin, fragment interference, recovery-interaction
model, and strength discretization.

The margin study of Appendix~\ref{sec:screening-margin} and the Grid
comparison of Appendix~\ref{sec:grid-enumeration} use all 7,321 requests.
Component ablations use a separate 1,111-request diagnostic cohort enriched
for challenging deployment settings, with an initial screening margin of
0.04. The cohort contains 195, 191, 279, 150, and 296 requests in S1 to S5.
In each scenario, it combines requests taken uniformly at random with requests
chosen by tags that mark informative cases, such as tight quality or latency
budgets, margin sensitivity, near-tied candidates, and multi-fragment or
geometric solutions, and it includes requests that contain Image editing. The
tags use request inputs and offline solver output only. Rates on the cohort
are pooled over its 1,111 requests, of which \sysfull{} satisfies 1,061
(95.50\%).

\subsection{Impact of the Screening Margin}
\label{sec:screening-margin}

\begin{table*}[!t]
\centering
\caption{Effect of the initial screening margin on request satisfaction and image fidelity.}
\label{tab:screening-margin}
\papertablestyle
\begin{adjustbox}{max width=\linewidth}
\begin{tabular}{@{}l c cccccc @{\hspace{6pt}} cc@{}}
\toprule
\toprule
& & \multicolumn{6}{c}{\textit{Request satisfaction (\%) $\uparrow$}} & \multicolumn{2}{c}{\textit{Image fidelity}} \\
\cmidrule(lr){3-8} \cmidrule(lr){9-10}
Method & $m_0$ & S1 & S2 & S3 & S4 & S5 & \textbf{Avg.} & PSNR $\uparrow$ & SSIM $\uparrow$ \\
\midrule
\sysgreedy{} & 0.04 & \underline{91.23} & 94.19 & 86.96 & 71.17 & 70.19 & \cellcolor{orange!8}82.75 & \textbf{41.77} & \textbf{0.9931} \\
\sysfull{} & 0.04 & \textbf{100.00} & \textbf{100.00} & \textbf{97.31} & \underline{82.24} & \textbf{98.34} & \cellcolor{orange!8}\underline{95.58} & \underline{41.61} & \textbf{0.9931} \\
\midrule
\sysgreedy{} & 0.06 & \textbf{100.00} & \underline{97.39} & \underline{87.55} & 76.03 & \underline{94.27} & \cellcolor{orange!8}91.05 & 40.91 & 0.9914 \\
\sysfull{} & 0.06 & \textbf{100.00} & \textbf{100.00} & \textbf{97.31} & \textbf{85.40} & \textbf{98.34} & \cellcolor{orange!8}\textbf{96.21} & 41.02 & \underline{0.9923} \\
\bottomrule
\bottomrule
\end{tabular}
\end{adjustbox}

\end{table*}

We evaluate how the initial rate-screening margin affects request coverage and image fidelity. We write $m_0$ for the initial value of the rate margin $\eta_{\mathrm{rate}}$ of Appendix~\ref{sec:method-acceptance-details}. Using the same 7,321 requests and fixed live-validation criteria, we compare $m_0=0.04$ and $m_0=0.06$. As shown in~\autoref{tab:screening-margin}, increasing the margin raises \sys{}'s scenario-averaged satisfaction from 95.58\% to 96.21\%, while mean PSNR over accepted requests decreases from 41.61 to 41.02~dB.

The coverage gain is concentrated in S4, where satisfaction increases from 82.24\% to 85.40\%, corresponding to 26 additional accepted requests among its 822 requests; the other four scenario rates remain unchanged. The PSNR columns of \autoref{tab:screening-margin} average over each setting's own accepted requests, so we also compare the two settings on the 7,104 requests that both accept. There, mean PSNR is 41.61~dB at $m_0=0.04$ and 41.05~dB at $m_0=0.06$, a paired decrease of 0.56~dB. The greedy variant is more sensitive to the same change: its satisfaction increases from 82.75\% to 91.05\%, and on the 5,922 requests that it accepts under both settings its mean PSNR decreases from 41.82 to 41.13~dB. A larger initial margin thus selects stronger configurations, which satisfy more of the difficult requests and lower fidelity on the requests that both settings accept, while the full search remains comparatively stable across the two settings.

\subsection{Continuous vs. Grid Strengths}
\label{sec:grid-enumeration}

We evaluate whether continuous strength optimization provides an advantage over selecting strengths from a fixed grid.

\begin{figure}[!htbp]
\centering
\includegraphics[width=\linewidth]{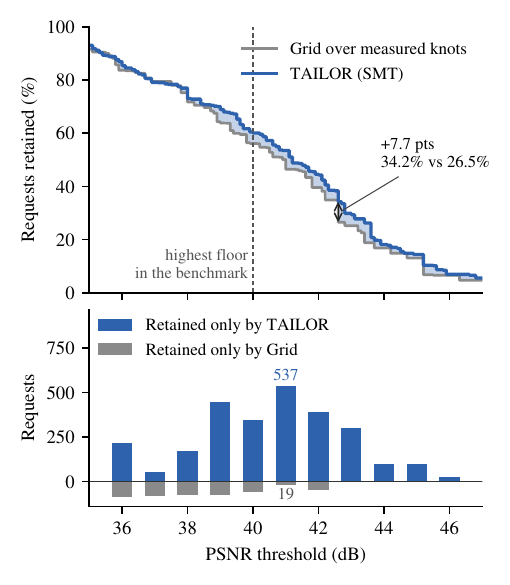}
\caption{\rev{Fidelity-threshold analysis.}}
\label{fig:fidelity-sweep}
\end{figure}

\begin{revblock}
\paragraph{Grid Enumeration.}
We compare \sys{} on all 7,321 requests with an enumeration baseline,
Grid. Grid enumerates the measured strength knots of VINE, TrustMark, and
VideoSeal (Table~\ref{tab:native-strength}): all nonempty subsets, all
embedding orders, all compatible geometric stages, and the Cartesian product of
knots, 58,433 configurations in total. The enumeration is exhaustive over
these discrete candidates at selection time. It is not an exhaustive live
measurement, since only the best structures are validated live. For each
selected fragment, Grid reads the mean recovery, the acceptance rate, and the
standalone distortion measured at the knot. Where a quantity was profiled on a
coarser grid that lacks the knot, as for Image editing, UnMarker, and some
stage responses, Grid evaluates the stored curve of \sys{} at the knot
instead. Recovery interactions and the pairwise distortion correction use the
same curves as \sys{}. Every configuration is checked
with the same path count $T(c)$, threshold, rate gate, contribution constraint,
and initial margin as \sys{}. Grid ranks the feasible configurations by
predicted distortion, keeps the three best structures at each margin level,
and passes them through the same live validation and margin fallback as
\sys{}.

\paragraph{Fidelity-Threshold Analysis.}
Figure~\ref{fig:fidelity-sweep} compares the fidelity of frozen
configurations selected by \sys{} and Grid Enumeration. Above 37.7~dB,
\sys{} retains at least as many accepted requests as Grid. At 40~dB, \sys{}
retains 60.1\% of requests compared with 56.2\% for Grid. The largest
difference occurs at 42.6~dB, where \sys{} achieves 34.2\% versus 26.5\%, a
gain of 7.7 percentage points. Configurations are not re-optimized for these
thresholds.
\end{revblock}

\subsection{Direct Fragment Superposition}
\label{sec:fragment-superposition}

We evaluate whether VINE, TrustMark, and VideoSeal remain independently decodable when embedded together, since complementary robustness is useful only if the fragments can coexist with limited interference. We compare standalone and sequentially composed watermarks on the same 100 $512\times512$ images from the evaluation pool, containing 25, 25, 25, 15, and 10 images from the five source groups, respectively. We use the deployed VINE-R, TrustMark-B, and VideoSeal implementations with all embedding strengths fixed to one. The experiment covers all three standalone fragments, six ordered pairs, and six triple permutations, for a total of 1,500 configuration-image measurements and 3,300 selected-fragment readouts. For each image, the same 100-bit codeword and fragment-specific key transformations are used across configurations. No attack, score fusion, or geometric recovery is applied, and every selected fragment is decoded independently from the composed image.

\begin{table*}[!t]
\centering
\caption{Independent raw-bit decoding after watermark superposition on 100 images.}
\label{tab:fragment-superposition}
\papertablestyle
\begin{tabular}{@{}l ccc c@{}}
\toprule
\toprule
& \multicolumn{3}{c}{\textit{Bit accuracy (\%)}} & \textit{All selected fragments (\%)} \\
\cmidrule(lr){2-4}\cmidrule(lr){5-5}
Embedding order & VINE & TrustMark & VideoSeal & $\geq64/100$ bits \\
\midrule
$V$ & 99.97 & - & - & 100 \\
$T$ & - & 100.00 & - & 100 \\
$S$ & - & - & 99.59 & 100 \\
\midrule
$V\!\to\!T$ & 99.98 & 100.00 & - & 100 \\
$V\!\to\!S$ & 99.96 & - & 99.48 & 100 \\
$T\!\to\!V$ & 99.97 & 100.00 & - & 100 \\
$T\!\to\!S$ & - & 99.96 & 99.58 & 100 \\
$S\!\to\!V$ & 99.96 & - & 99.21 & 100 \\
$S\!\to\!T$ & - & 99.98 & 96.95 & 100 \\
\midrule
$V\!\to\!T\!\to\!S$ & 99.95 & 99.97 & 99.52 & 100 \\
$V\!\to\!S\!\to\!T$ & 99.95 & 99.98 & 96.78 & 100 \\
$T\!\to\!V\!\to\!S$ & 99.96 & 99.96 & 99.51 & 100 \\
$T\!\to\!S\!\to\!V$ & 99.97 & 99.96 & 99.27 & 100 \\
$S\!\to\!V\!\to\!T$ & 99.96 & 99.99 & 95.54 & 100 \\
$S\!\to\!T\!\to\!V$ & 99.97 & 99.98 & 95.72 & 100 \\
\bottomrule
\bottomrule
\end{tabular}
\end{table*}

In Table~\ref{tab:fragment-superposition}, V, T, and S denote VINE, TrustMark, and VideoSeal, and arrows indicate embedding order. Bit accuracy is computed from the signs of the aligned logits against the expected codeword, with zero values counted as errors. Each mean is computed over all 100 images without filtering. We use a fixed diagnostic threshold of 64 matching bits out of 100 for the individual fragment readouts. This common threshold enables consistent comparisons of interference across embedding orders. Deployment verification instead uses the configuration-dependent threshold defined in Eq.~\ref{eq:unified-threshold}.

Standalone mean bit accuracy is 99.97\% for VINE, 100.00\% for TrustMark, and 99.59\% for VideoSeal. Across all ordered pairs and triples, the corresponding composed readouts average 99.37\%, compared with 99.85\% for their standalone counterparts, a decrease of 0.49 percentage points. The largest mean loss is 0.02 points for VINE and 0.04 points for TrustMark. VideoSeal is more order-sensitive: its loss is at most 0.08 points when it is embedded last in a triple and reaches 4.05 points under other orders. In this clean-image setting, every readout of every tested composition exceeds the diagnostic threshold on all 100 images.

At unit strength and without attacks, the three fragments therefore coexist with little loss at this diagnostic threshold, while embedding order matters, particularly for VideoSeal. Together with their complementary robustness across attack scenarios, this supports VINE, TrustMark, and VideoSeal as compatible building blocks for the fragment library of \sys{}.

\subsection{Recovery-Interaction Sensitivity}
\label{sec:recovery-interaction-ablation}

We further test how sensitive \sys{} is to explicit modeling of recovery
interactions between composed fragments. The NoRecoveryInteraction variant
sets every pairwise recovery correction $\delta_{g\to f}$ to zero and
replaces the stage responses that were measured with a second fragment
embedded afterwards by the corresponding single-fragment stage responses
(Appendix~\ref{sec:method-table-details}). The distortion model, including
$e_{fg}$ and the stage term $D_\varphi$, is unchanged, and live validation,
local refinement, and margin fallback are retained. The variant achieves 95.41\% request satisfaction on the ablation cohort,
compared with 95.50\% for \sysfull{}. The marginal difference indicates that
explicit recovery-interaction corrections have limited impact on the
evaluated fragment library.

\section{Qualitative Request Examples}
\label{sec:qualitative-requests}

We provide qualitative examples to illustrate both how \sys{} compares with
fixed-strength controls and how its selected composition differs between
deployment requests.

Figures~\ref{fig:qualitative-s2} to \ref{fig:qualitative-s4} compare \sys{}
with fixed-strength standalone watermarks and Enumeration across
representative deployment requests.
\begingroup
\makeatletter
\setlength{\@fptop}{0pt}
\setlength{\@dblfptop}{0pt}
\makeatother

\begin{figure*}[t]
\centering
\includegraphics[width=\linewidth]{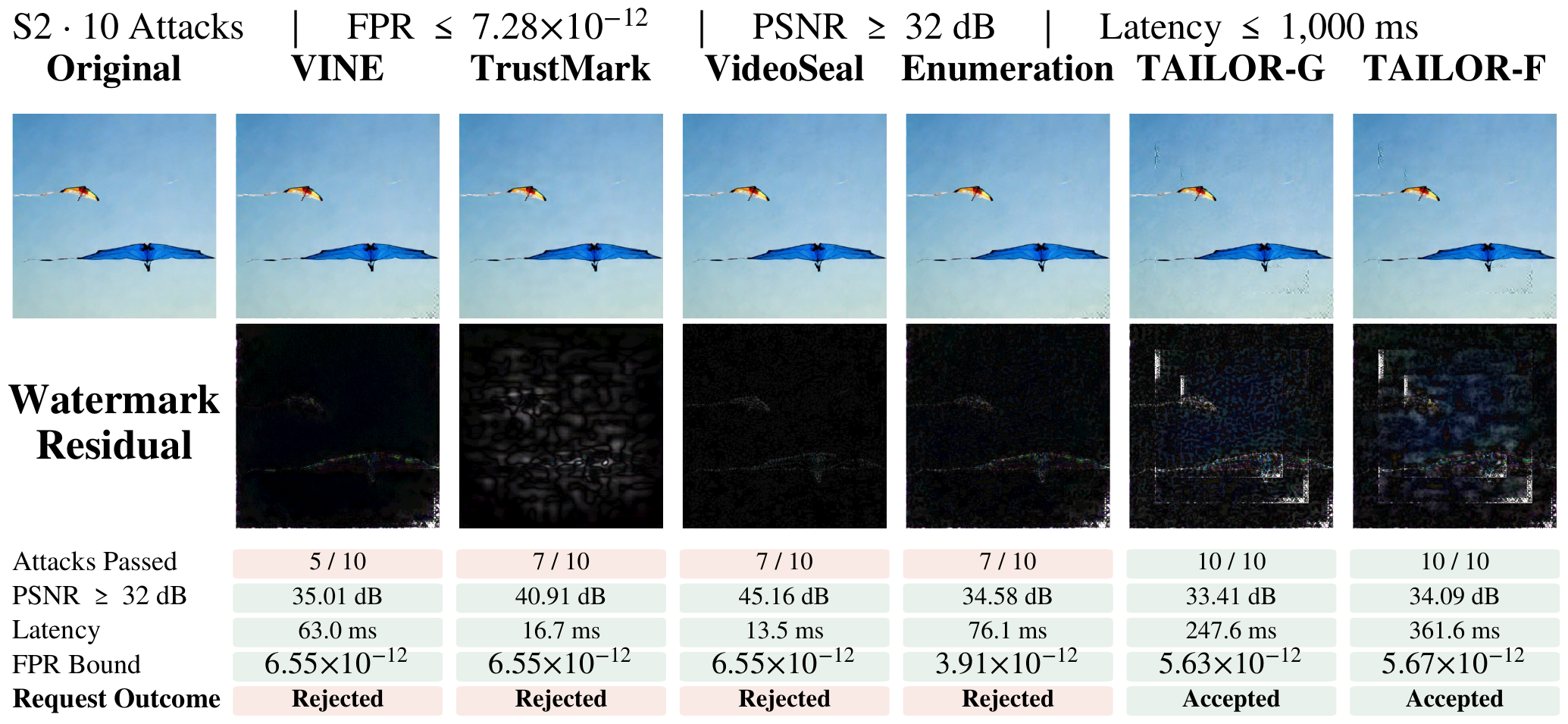}
\caption{Watermarked images, residuals, and constraint outcomes for a geometric-transformation request (S2). The image shown is one of the 100 live images, PSNR and latency are computed over the live images, and FPR Bound is the analytical bound of Eq.~\ref{eq:fpr-accounting}, not a measured rate. Enumeration is a supplementary fixed-strength control.}
\label{fig:qualitative-s2}
\end{figure*}

\begin{figure*}[t]
\centering
\includegraphics[width=\linewidth]{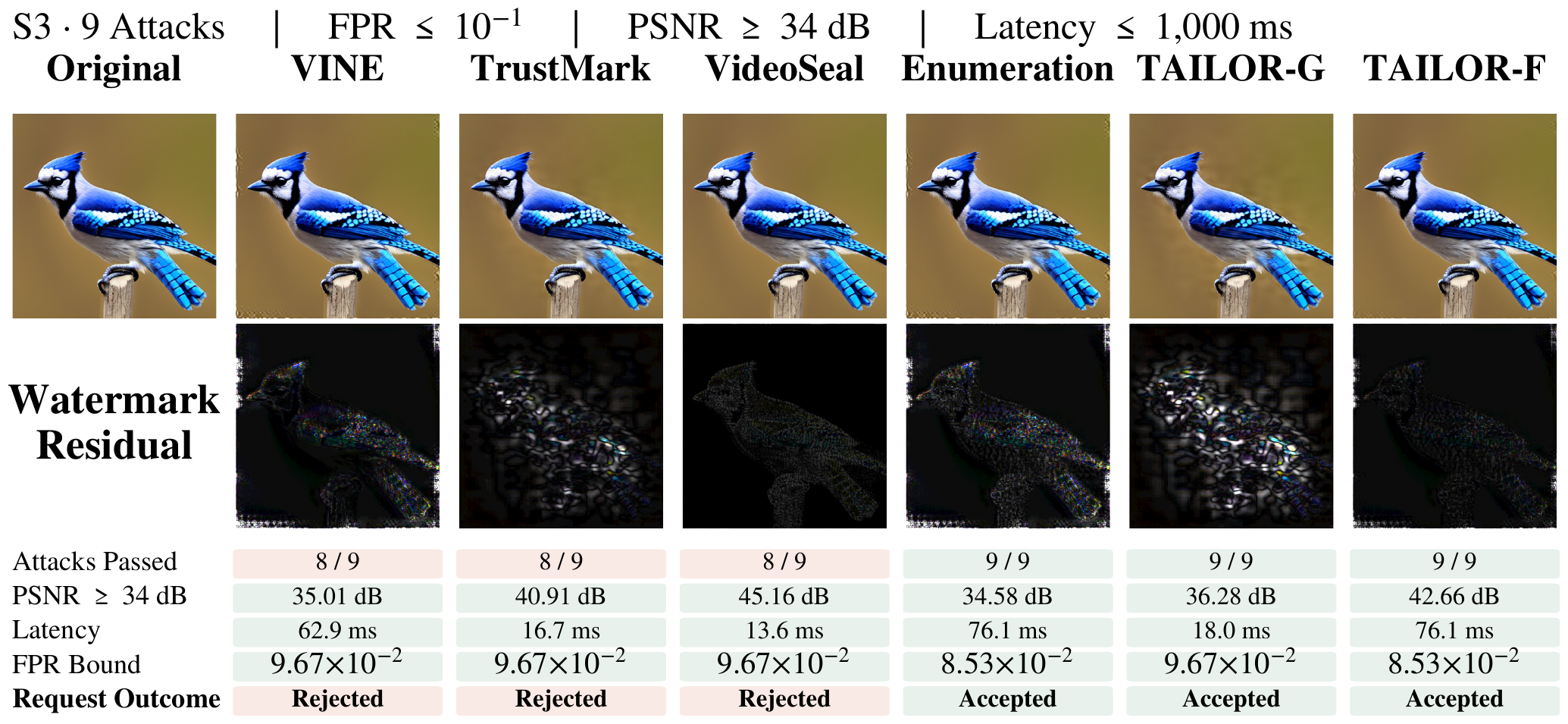}
\caption{Watermarked images, residuals, and constraint outcomes for a regeneration-and-editing request (S3). Conventions as in Figure~\ref{fig:qualitative-s2}.}
\label{fig:qualitative-s3}
\end{figure*}

\begin{figure*}[t]
\centering
\includegraphics[width=\linewidth]{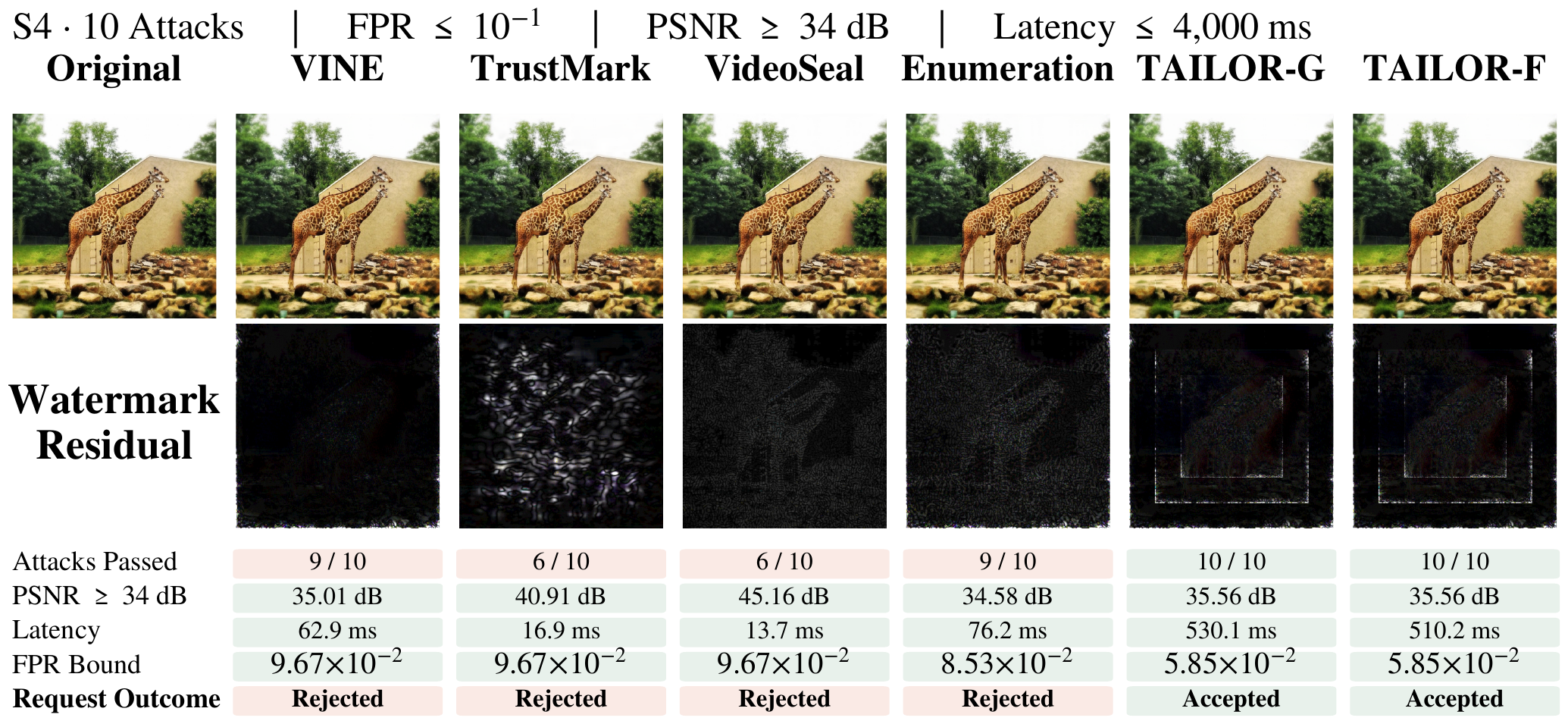}
\caption{Watermarked images, residuals, and constraint outcomes for a watermark-removal request (S4). Conventions as in Figure~\ref{fig:qualitative-s2}.}
\label{fig:qualitative-s4}
\end{figure*}
\endgroup

Figures~\ref{fig:request-solution-s1}, \ref{fig:request-solution-s3}, and
\ref{fig:request-solution-s5} trace three further requests from specification
to the configuration accepted by \sys{}. Each figure shows the requested
attacks and budgets, the selected fragments and recovery stage, and the
corresponding live-validation results. For multi-fragment configurations, the
columns under Individual Fragment Readouts report the pass rate of each
fragment's own readout in the composed image at the configuration threshold,
read on the view that the verifier adopted for the image. The column under
Full Verifier reports the pass rate of the complete decoder, which also uses
the fused readout and the geometric views. These columns illustrate how
different fragments contribute coverage to different attacks. The three
examples show a single-fragment, a two-fragment, and a three-fragment
configuration under different deployment requirements: a single TrustMark
fragment, a VideoSeal-VINE composition, and all three fragments with scale
search. The requests differ in their attack sets and in all three budgets, so
the examples are illustrations and not a controlled comparison.

In all six figures, the image shown is one of the 100 live images, whereas the
reported PSNR, pass rates, and latency are computed over the live images: PSNR
is the mean embedding PSNR, a pass rate is the fraction of images accepted
under an attack, and latency is the measured value of
Eq.~\ref{eq:deployment-latency}. The FPR bound is the analytical bound of
Eq.~\ref{eq:fpr-accounting} for the deployed configuration and not a measured
false-positive rate. Residuals use the same $12\times$ display gain.

\begingroup
\makeatletter
\setlength{\@fptop}{0pt}
\setlength{\@dblfptop}{0pt}
\makeatother

\begin{figure*}[t]
\centering
\includegraphics[width=\linewidth]{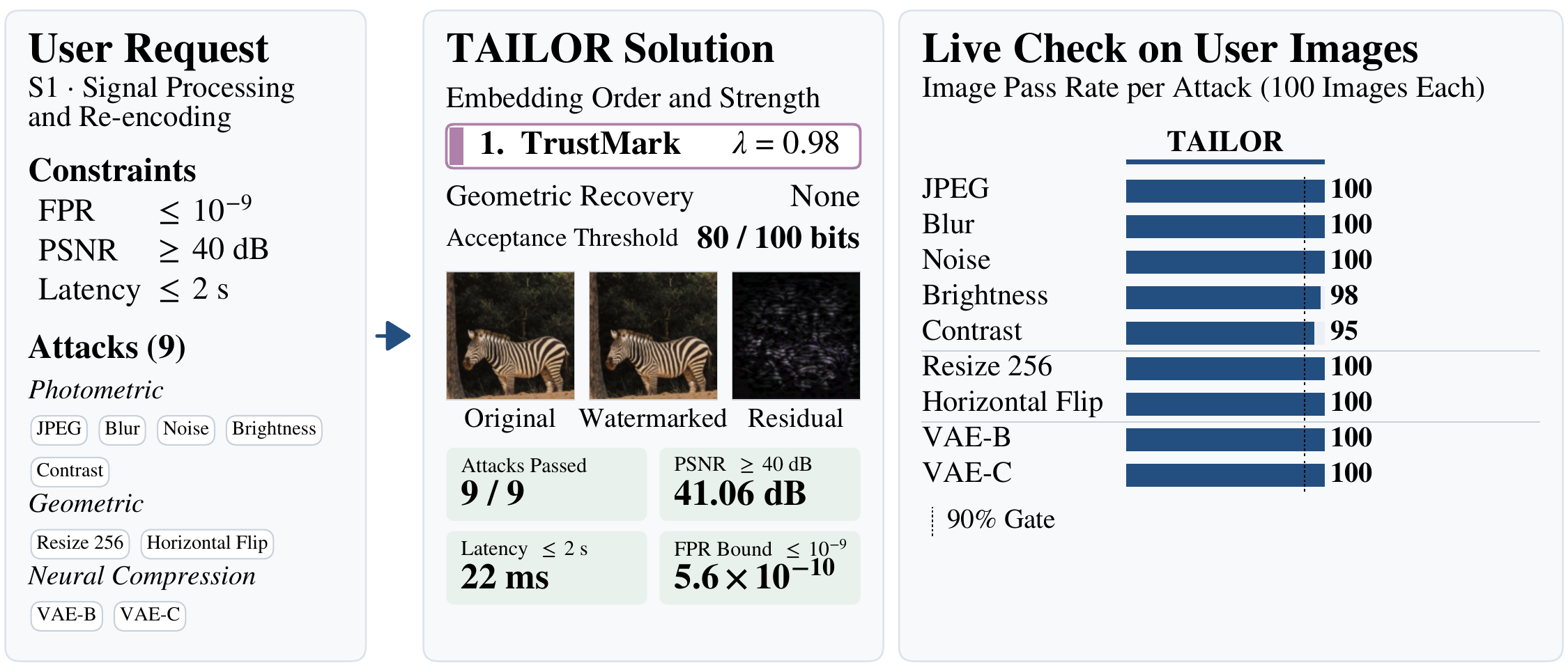}
\caption{A signal-processing request (S1) answered by a single fragment. The image shown is one of the 100 live images. Pass rates, PSNR, and latency are computed over the live images, and FPR Bound is the analytical bound of Eq.~\ref{eq:fpr-accounting}, not a measured rate.}
\label{fig:request-solution-s1}
\end{figure*}

\begin{figure*}[t]
\centering
\includegraphics[width=\linewidth]{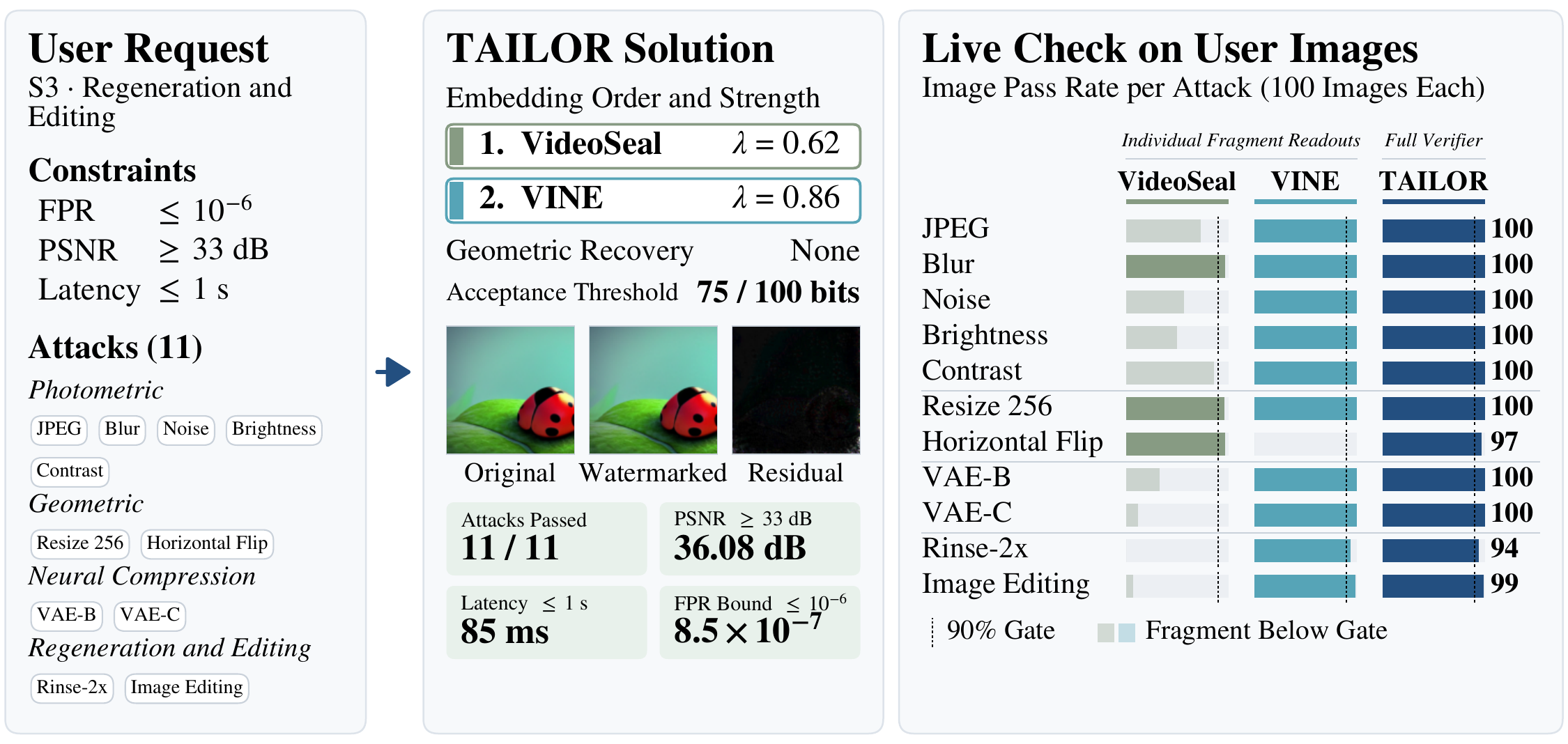}
\caption{A regeneration-and-editing request (S3) answered by a complementary fragment pair. Individual Fragment Readouts: pass rate of each fragment's own readout in the composed image at the configuration threshold. Full Verifier: pass rate of the complete decoder. Other conventions as in Figure~\ref{fig:request-solution-s1}.}
\label{fig:request-solution-s3}
\end{figure*}

\begin{figure*}[t]
\centering
\includegraphics[width=\linewidth]{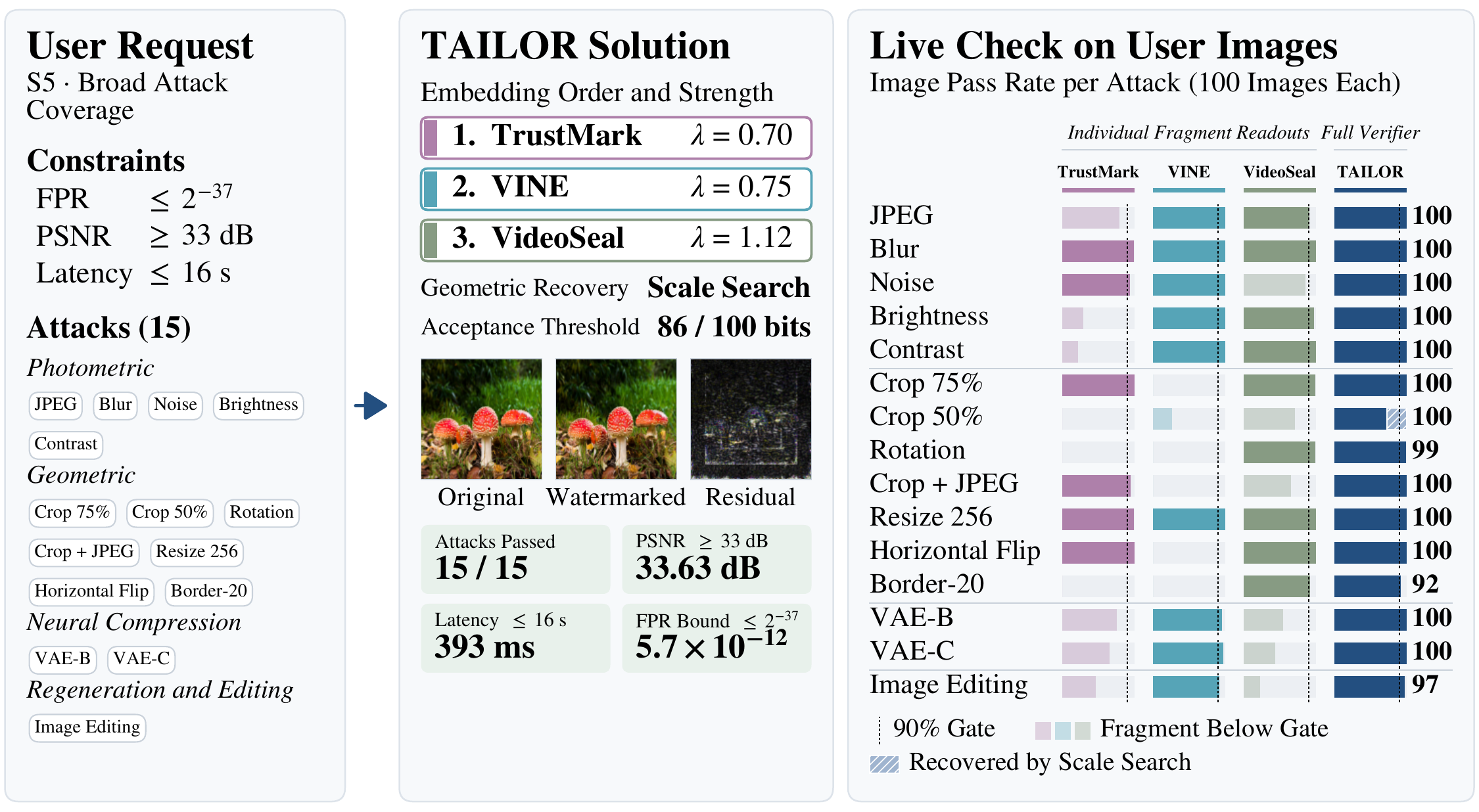}
\caption{A broad-coverage request (S5) answered by three fragments with scale recovery. Conventions as in Figures~\ref{fig:request-solution-s1} and~\ref{fig:request-solution-s3}. Individual readouts are taken on the view that the verifier adopted for each image.}
\label{fig:request-solution-s5}
\end{figure*}
\endgroup

\end{document}